\documentclass[a4paper,11pt]{article}
\usepackage[top=3cm,bottom=3cm]
{geometry}
\usepackage{amsmath,amsthm,amssymb,xspace,xcolor,hyperref,bm,times,graphicx}
\hypersetup{colorlinks=true, linkcolor=black,citecolor=black,urlcolor=black}
\numberwithin{equation}{section}

\definecolor{mysil}{rgb}{0.6,0.6,0.6}

\theoremstyle{plain}
\newtheorem{theorem}{Theorem}

\newtheorem{proposition}{Proposition}
\newtheorem{lemma}{Lemma}

\newtheorem{definition}{Definition}

\newtheorem{problem}{Problem}
\newtheorem{remark}{Remark}

\newcommand{\APC}{\ensuremath{{\sf APC_1}}\xspace}

\newcommand{\SB}{\ensuremath{{\sf S^1_2}}\xspace}

\newcommand{\PV}{\ensuremath{{\sf PV_1}}\xspace}

\newcommand{\ZFC}{\ensuremath{{\sf ZFC}}\xspace}

\newcommand{\EF}{\ensuremath{{\sf EF}}\xspace}

\newcommand{\CP}{\ensuremath{{\sf CP}}\xspace}
\newcommand{\CF}{\ensuremath{{\sf CF}}\xspace}
\newcommand{\Res}{\ensuremath{{\sf Res}}\xspace}
\newcommand{\SYS}{\ensuremath{{\sf SYS}}\xspace}
\newcommand{\IPS}{\ensuremath{{\sf IPS}}\xspace} 
\newcommand{\Gs}{\ensuremath{{\sf G}_1^*}\xspace}

\newcommand{\VP}{\ensuremath{{\sf VP}}\xspace}
\newcommand{\VNP}{\ensuremath{{\sf VNP}}\xspace}

\newcommand{\Ptime}{\ensuremath{{\sf P}}\xspace}
\newcommand{\Ppoly}{\ensuremath{{\sf P/poly}}\xspace}
\newcommand{\NC}{\ensuremath{{\sf NC}^1}\xspace}
\newcommand{\AC}{\ensuremath{{\sf AC}^0}\xspace}

\newcommand{\NP}{\ensuremath{{\sf NP}}\xspace}
\newcommand{\coNP}{\ensuremath{{\sf coNP}}\xspace}

\newcommand{\Etime}{\ensuremath{{\sf E}}\xspace}

\newcommand{\Circuit}{\ensuremath{{\sf Circuit}}\xspace}

\newcommand{\SAT}{\ensuremath{{\sf SAT}}\xspace}

\newcommand{\MCSP}{\ensuremath{{\sf MCSP}}\xspace}

\newcommand{\ttable}{\ensuremath{{\sf tt}}\xspace}
\newcommand{\lb}{\ensuremath{{\sf lb}}\xspace}
\newcommand{\PHP}{\ensuremath{{\sf PHP}}\xspace}

\newcommand{\colone}{\mathbin{\vcenter{\hbox{$:$}}\hspace{-3pt}=}}

\begin{document}

\title{{\bf \textsc{Efficient Adversaries}}}
\author{{\fontfamily{ppl}\selectfont Erfan Khaniki\footnotemark[2]}
\and {\fontfamily{ppl}\selectfont J\'an Pich\footnotemark[2]} 
\and {\fontfamily{ppl}\selectfont Dmitry Sokolov\footnotemark[3]}
}
\date{\small November 2026\footnote{A preliminary version of this paper appeared in Computational Complexity Conference 2026.}}
\maketitle

\renewcommand{\thefootnote}{\fnsymbol{footnote}}
\footnotetext[2]{University of Oxford}
\footnotetext[3]{Universit{\'{e}} de Montr{\'{e}}al}
\renewcommand{\thefootnote}{\arabic{footnote}}

\begin{abstract} 
The size of Frege proofs can be characterized in terms of prover-adversary games of Pudl\'ak and Buss. We consider a generalization of prover-adversary games to many standard proof systems and show that some of the major proof complexity lower bounds such as the constant-depth Frege lower bound for the pigeonhole principle based on the method of $k$-evaluations, the Resolution lower bound for the weak pigeonhole principle based on the method of pseudo-width and Razborov's $\Res(k)$ lower bound for Nisan-Wigderson generators based on expansion and a width lower bound (which is used to derive the $\Res(k)$-hardness of formulas expressing circuit lower bounds)
 are constructive in the sense that they yield efficient algorithms computing winning strategies of adversaries in the generalized games. This is in contrast~with our second result saying that if $(a)$ such a constructive lower bound exists for Extended Frege system \EF for formulas expressing succinct circuit lower bounds for \SAT and $(b)$ \EF is strong enough to prove efficiently the correctness of anticheckers for \SAT, then it is easy to separate the canonical pair of \EF.
\end{abstract}

\section{Introduction}
Proving circuit lower bounds for explicit Boolean functions is a central open problem in complexity theory, underlying questions such as the P vs NP problem.

The notorious difficulty of the problem has led to the development of a theory of formal barrier results. For example, already in the 80's Razborov \cite{Ram} showed that the approximation method, used to derive monotone circuit lower bounds \cite{Rmlb} and lower bounds against $\AC[p]$ \cite{Rac2,Sac2}, cannot yield superquadratic lower bounds for non-monotone circuits. In the 90s, Razborov and Rudich \cite{Rnat} discovered the influential natural proofs barrier clarifying the limitations of major lower bound methods such as the method of random restrictions or the algebraic method of Razborov and Smolensky \cite{Rac2,Sac2}. More recently, the locality barrier explained why a successful implementation of hardness magnification, an approach to strong explicit circuit lower bounds sidestepping the natural proofs barrier, requires more substantial new ideas \cite{CHOPRS}. Such barrier results could potentially guide us towards the solution of the original problem, but it is also possible that they will eventually lead to the conclusion that strong circuit lower bounds are practically impossible to establish or to the discovery of new surprising algorithms, cf. \cite{CIKK}.
\smallskip

The possibility that circuit lower bounds could lie beyond what can be achieved within feasible logical reasoning was investigated extensively by Razborov who used natural proofs to derive the unprovability of circuit lower bounds in certain theories of bounded arithmetic \cite{Rba}. Such unprovability results can be typically interpreted as lower bounds on lengths of proofs of tautologies expressing circuit lower bounds in propositional proof systems. Indeed, after composing \cite{Rba} Razborov switched to the more elegant and transparent framework of propositional logic and formulated his subsequent results in this setting; including a degree lower bound for deriving circuit lower bounds in Polynomial Calculus \cite{Rpc}, a Resolution lower bound for formulas expressing circuit lower bounds based on the pseudo-width method \cite{Rpw1,Rpw2,Rpw3} (which simplified and strengthened the original result of Raz \cite{Rwphp}), and the hardness of circuit lower bounds for $\Res(k)$, an extension of Resolution operating with $k$-DNFs, for a small parameter $k$ \cite{Rkdnf}. The last result remains the strongest propositional lower bound for tautologies encoding circuit lower bounds.\footnote{There are several related results about the impossibility of deriving strong circuit lower bounds in bounded arithmetic \cite{Pclba,PSapc,LO,CLOlmcs,tsin} which appear to be incomparable to the lower bounds for propositional systems, but it seems that strengthening them in a more substantial way requires new propositional lower bounds. In the propositional setting, Austrin and Risse \cite{AR} showed that Sum-of-Squares proof system, which proves efficiently the pigeonhole principle, cannot prove efficiently circuit lower bounds for certain functions $f$, but the lower bound does not hold for all functions $f$. An analogous meta-mathematical theory is currently under development also in the algebraic setting \cite{LST}.} 

Ultimately, in \cite{Rkdnf} Razborov formulated intriguing conjectures about a conditional hardness of Nisan-Wigderson (NW) generators for Frege and Extended Frege system \EF: suitable NW-generators should form good proof complexity generators in the sense that tautologies postulating the existence of elements outside their range require super-polynomial-size proofs in Frege systems, assuming \Ptime is hard on average for \NC, and in \EF, assuming $\NP\cap \coNP$ is hard on average for \Ppoly. If true, these conjectures would present a dramatic barrier establishing that feasibly-constructive logical reasoning (which can be defined in terms of short \EF proofs or as the provability in Cook's theory \PV \cite{Kba,Cpv}) cannot prove explicit circuit lower bounds --- \Ppoly lower bounds in the case of Frege and lower bounds for a specific kind of $(\NP\cap\coNP){\sf/poly}$ circuits in the case of \EF.\footnote{A lot of contemporary complexity theory is formalizable within \EF, e.g., the PCP theorem has short \EF proofs \cite{Ppcp}, see also \cite{Igorsurvey}. Some candidates for \PV-unprovable theorems from computational complexity are given in \cite{CLO} but they exploit witnessing theorems which (under standard hardness assumptions) cannot be used to get the \PV-unprovability of deterministic circuit lower bounds for \SAT \cite{Pclba,MP}.} A similar earlier conjecture of Kraj\'{i}\v{c}ek \cite[Conjecture 7.9]{Kwphp}, implying that, under the existence of strong pseudorandom generators, subexponential  circuit lower bounds are hard for \EF was motivated by the role of dual weak pigeonhole principle in bounded arithmetic. In \cite{Kpcg}, Kraj\'i\v{c}ek formulated the conjecture in an unconditional form.

The conjectures of Razborov and Kraj\'i\v{c}ek are interesting not only because of the potential meta-mathematical consequences but also because they suggest the existence of a strong connection between circuit complexity and proof complexity. 
\medskip

Proof complexity presents a logic-oriented approach to the separation of \NP and \coNP which proceeds by proving lower bounds on lengths of proofs in increasingly more powerful propositional proof systems. 
Unfortunately, Razborov's unprovability results as well as other existing proof complexity lower bounds work only for very weak proof systems. In fact, in some sense proving proof complexity lower bounds has turned out to be harder than proving circuit lower bounds: turning \AC lower bounds into constant-depth Frege lower bounds required developing a substantially more involved adaptation of the method of random restrictions in the form of $k$-evaluations \cite{Aphp,KPW,PBI} and it still remains an open problem to derive lower bounds for Frege systems operating with $\AC[2]$ circuits despite the existence of explicit $\AC[2]$ lower bounds \cite{Rac2,Sac2}. From the perspective of the attempts to derive propositional lower bounds for tautologies expressing circuit lower bounds this is somewhat ironic as it suggests that deriving such hardness results unconditionally might be harder than proving circuit lower bounds themselves.

Generally, the relation between circuit complexity and proof complexity is rather mysterious. Formal connections between the two areas exist if we restrict ourselves to weak computational models. For example, the method of feasible interpolation derives proof complexity lower bounds for many weak proof systems from circuit lower bounds for monotone circuits and related models \cite{Kpc}. In the opposite direction, lifting theorems yield lower bounds for various monotone computational models from lower bounds for weak propositional proof systems, see e.g. \cite{Rth,Gth}. In the algebraic setting, \IPS lower bounds imply $\VP\ne\VNP$ \cite{GP}. Unfortunately, these connections break when moving to strong proof systems and the Boolean setting. Recently a conditional connection between lower bounds for strong proof systems and strong circuit lower bounds such as $\NP\not\subseteq\Ppoly$ has been established \cite{PSeopt,AKPS}, but it requires the assumption that the proof system proves efficiently some circuit lower bound such as ${\sf E}$ being hard on average for subexponential-size circuits.

As a consequence of the lack of formal links between circuit and proof complexity, the barrier results in circuit complexity do not seem to say much about the possibility of deriving proof complexity lower bounds for strong proof systems. The question of identifying a ``natural proofs"-like barrier in proof complexity was raised already by Beame and  Pitassi \cite{BP}. Given that some prominent proof complexity lower bounds are harder to derive than their counterparts in circuit complexity, it might seem even more ironic that we have not been able to identify such a barrier, see Section \ref{s:related} for a discussion of earlier attempts. Is there an explanation for our inability to prove Frege lower bounds like those in Razborov's conjecture? In this paper we introduce a notion that has a potential to answer this question. 

\subsection{Our contribution}

Following the paradigm of natural proofs, which turn constructive circuit lower bounds into efficient algorithms, we want to find a way to interpret (sufficiently constructive) proof complexity lower bounds as a construction of new surprising algorithms.

In order to get closer to such a conclusion we focus on prover-adversary games 
used by Pudl\'ak and Buss \cite{PB} 
to characterize the lengths of proofs in Frege systems. 
We outline a slightly modified formulation of these games which is sufficiently general to capture uniformly the existing lower bounds for many different proof systems.
\medskip

An s-size lower bound for a formula $\phi$ in a proof system $P$ is often obtained by defining a strategy for an adversary who responds to queries of the prover in the following manner. Given a supposed $s$-size $P$-refutation $\pi$ of $\neg\phi$, the prover sends to the adversary the last formula $\psi$ of $\pi$ and the adversary responds with 1 or 0. 
After that, the prover sends formulas $\psi_1,\dots,\psi_k$ from which $\psi$ was derived in $\pi$ by a single application of a derivation rule and the adversary responds with 1 or 0 to each of them. The prover then proceeds to a subproof of $\pi$ which derives a formula $\psi_i\in \{\psi_1,\dots,\psi_k\}$ assigned 0 by the adversary, if such a formula exists. If such a formula does not exist and the adversary assigned 0 to $\psi$, the adversary has been proven to be inconsistent (caught). The game continues in this way possibly until the players reach the axioms or $\neg \phi$. The adversary has to assign 0 to the contradiction (which is the last formula $\psi$, if $\pi$ is a refutation) and 1 to the axioms and $\neg\phi$ (indicating that the adversary posses an assignment falsifying $\phi$). 
If we can show that the adversary maintains consistency, we can conclude that $\pi$ was not a correct refutation of $\neg\phi$. For a more elaborated discussion of the prover-adversary games see Section \ref{s:algfromadv}.

Intuitively, we define a constructive proof complexity lower bound by the existence of an ``efficient adversary"; more precisely, by an efficient algorithm generating the responses of the adversary on given formulas such that for each short $P$-refutation (not necessarily of $\neg\phi$) the adversary (who might not be given the entire refutation) stays consistent in the game described above, cf. Definitions \ref{d:ead}-\ref{d:evalead}. The adversary always has to assign 1 to $\neg\phi$ and the axioms of $P$ but not necessarily to other initial formulas of a given $P$-refutation. A randomized efficient adversary is allowed to use randomness so that for each short $P$-refutation the adversary can choose a random string and use it to determine its responses. 
To indicate that an adversary corresponds to a lower bound (for a formula $\phi$) in a proof system $P$ we speak of $P$-adversaries (claiming that $\neg\phi$ is satisfiable).

Note that in order to make a proof complexity lower bound for a formula $\phi$ constructive, we do not need to construct adversaries for short refutations of $\neg\phi$ --- such refutations do not exist if the lower bound holds. In a sense, an efficient adversary rules out more than just the existence of a short refutation of $\neg\phi$, it could be said that it rules out the existence of a short `pseudoproof' of $\phi$, cf. Proposition \ref{p:fadv}.
\medskip

Our first theorem shows that constructive \EF lower bounds for formulas $\lb_A(\SAT,n^2)$ (expressing that \SAT is not computable by $n^2$-size circuits on a set of inputs $A\subseteq\{0,1\}^n$) given by randomized efficient \EF-adversaries\footnote{The definition of \EF-adversaries requires a careful treatment of the extension axioms, see Section \ref{s:algfromadv}.} claiming that $\neg \lb_A(\SAT,n^2)$ is satisfiable can be used to break the canonical pair of \EF assuming that \EF is capable of efficiently deriving a statement about anticheckers. Informally, the canonical pair of a proof system $P$ consists of two disjoint \NP sets: the set of all satisfiable formulas and the set of all unsatisfiable formulas whose unsatisfiability has a short $P$-proof. The canonical pair of $P$ is hard to separate if there is no small circuit accepting all inputs from the first set and rejecting all inputs from the second set, cf. \cite{Rpairs}.

\begin{theorem}[From adversaries to algorithms -- Informal, cf. Theorem \ref{t:efadv}]\label{t:efadv-inf}\ \newline 
Assume 1 - 2: 
\begin{itemize}
\item[1.] {\em (The canonical pair of \EF is hard.
)}
For each $poly(n)$-size circuit $D$ with $n$ inputs, there is $z$ such that $D(z)=0$ but $z$ is a satisfiable formula or $D(z)=1$ but there is a $poly(n)$-size \EF-refutation of $z$. 
\item[2.] {\em (Provably constructive anticheckers.)} There is $A\subseteq\{0,1\}^n$, $|A|=poly(n)$, 
and a $poly(n)$-size circuit $F$ such that for each $z\in\{0,1\}^n$ there is a $poly(n)$-size \EF-proof of the statement expressing that ``if an $n^2$-size circuit $B$ solves \SAT on $A$, then a $poly(n)$-size circuit $F(B)$ solves \SAT on $z$". 
\end{itemize}
Then for each randomized $poly(n)$-time \EF-adversary claiming that $\neg\lb_A(\SAT,n^2)$ is satisfiable, there is a $poly(n)$-size \EF-refutation (not necessarily of $\neg\lb_A(\SAT,n^2)$) catching the adversary 
with probability $\ge 1/2-1/n$ over the randomness of the adversary.
\end{theorem}


Reading it contrapositively, Theorem \ref{t:efadv-inf} says that efficient adversaries defining constructive \EF lower bounds for formulas $\lb_A(\SAT,n^2)$ yield efficient algorithms breaking the canonical pair of \EF (a condition similar to the weak automatability of \EF), assuming Condition 2 holds. Condition 2, the second assumption of the theorem, is less standard. It postulates the existence of a set of `anticheckers' $A$ such that a small circuit which solves \SAT on $A$ can be turned into a circuit which solves \SAT on any other input. Such $A$ is known to exist unconditionally, cf. Section \ref{s:algfromadv}, but Condition 2 additionally says that this fact is efficiently provable in \EF. It can be shown that Condition 2 follows if, e.g., \EF proves efficiently that ($a$) 
\Etime is hard on average for subexponential-size circuits and \EF proves efficiently that ($b$) the existence of one-way permutation can be based on $\NP\not\subseteq\Ppoly$. Both statements ($a$) and ($b$) are considered to be more approachable than non-uniform circuit lower bounds for \SAT. The plausibility of Condition 2 can be further enhanced by replacing \EF in Theorem \ref{t:efadv-inf} by stronger systems such as (a propositional equivalent of) \ZFC, the theorem remains valid in this case as well. As pointed out by Noel Arteche (private communication), Condition 2 implies also that $\NP\not\subseteq\Circuit[n^{\ell}]$, for each $\ell$. See Section \ref{s:algfromadv} for a more detailed discussion.

The parameter $1/2-1/n$ in Theorem \ref{t:efadv-inf} is not essential for the purpose of a barrier result, any constant would work equally well.

The proof of Theorem \ref{t:efadv-inf} is straightforward: Since the adversary claims that it has a circuit solving \SAT on $A$, using Condition 2 we can try to force it (with \EF-queries) to claim that it has a circuit $C$ which solves \SAT everywhere and to simulate the computation of circuit $C$ on any input $z$. This almost works except that we only get an efficient algorithm breaking the canonical pair of \EF: If the adversary claims that $C$ says that $z$ is a satisfiable formula and $z$ is unsatisfiable, we need the additional assumption that $z$ has a short \EF-refutation to reveal the inconsistency of the adversary (and force it to avoid such claims).

The 
proof works also for
formulas $\neg\ttable(\SAT,n^2)\colone\neg\lb_{\{0,1\}^n}(\SAT,n^2)$ instead of $\neg\lb_A$, but it crucially exploits the fact that 
the adversary is efficient w.r.t. the queries of the prover, not w.r.t. the formula for which the lower bound is being proved (which is of size $2^{O(n)}$ in this case). In fact, in the case of the \ttable-formulas we can avoid Condition~2 which can be replaced by a formalization of the search-to-decision reduction for \SAT. Further, the switch from $poly(n)$-size to $2^{O(n)}$-size formulas allows us to replace the first assumption in Theorem \ref{t:efadv-inf2} with $\SAT\notin\Ppoly$. As a result we get

\begin{theorem}[Informal, cf. Theorem \ref{t:efadv}]\label{t:efadv-inf2}\ \newline 
Assume that $\SAT\notin\Ppoly$. Then for each randomized p-time \EF-adversary claiming that $\neg\ttable(\SAT,n^2)$ is satisfiable, there is a $2^{O(n)}$-size \EF-refutation catching the adversary 
 with probability $\ge 1/2-1/n$ over the randomness of the adversary.
\end{theorem}



\medskip

\noindent {\bf Constructive proof complexity lower bounds.} We complement Theorem \ref{t:efadv-inf} by showing that some of the major existing proof complexity lower bounds are indeed constructive.
\medskip

As the first example we design efficient adversaries for the strongest known lower bound for $\lb$-tautologies due to Razborov \cite{Rkdnf}, who showed that there are no $2^{t^{\Omega(1)}}$-size $\Res(\Omega(\log t))$-refutations of $\neg\lb_W(f,t)$, for $n^2\le t\le 2^n$ and any $W\subseteq\{0,1\}^n$. More precisely, we design efficient adversaries for ``proof complexity generator" formulas $\tau_{W}(A,b)$ (expressing that a system of linear equations $Ax=b$ is solvable) instead of $\neg\lb_W(f,t)$. This is justified by the fact that Razborov's proof proceeds by reducing the lower bound for $\lb_W(f,t)$ to a lower bound for $\neg\tau_{W}(A,b)$ and Theorem \ref{t:efadv-inf} holds even if the formula $\neg\lb_W(\SAT,n^2)$ is replaced by any formula $\phi$ such that \EF proves efficiently $\phi\rightarrow\neg\lb_W(\SAT,n^2)$, cf. Remark \ref{r:phi}. The focus on the $\tau$-formulas allows us to avoid the problem of encoding the reduction of lower bounds for $\lb_W(f,t)$ to lower bounds for $\neg\tau_{W}(A,b)$ in the strategy of the adversary, see Section \ref{s:width} for more details.

\begin{theorem}[Constructive $\Res(k)$ lower bound of Razborov -- Informal, cf. Theorem \ref{t:efadv-r}]\label{t:efadv-r-inf}\ \newline 
For each sufficiently small $\delta>0$, there is $\epsilon>0$ and a $poly(t^{t^{\delta}})$-size randomized $\Res(k)$-adversary $C$ claiming that $\tau_{W}(A,b)$ is satisfiable, where $k=\epsilon\log t$ and $|W|=poly(t)$, such that for each $2^{t^{\Omega(\delta)}}$-size $\Res(k)$-refutation $\pi$ of any set of clauses, with probability $\ge 9/10$ over the randomness of $C$, $\pi$ does not catch $C$.
\end{theorem}

The complexity of the adversary in Theorem \ref{t:efadv-r-inf} is subexponential, but the formal version of Theorem \ref{t:efadv-inf} (Theorem \ref{t:efadv}) captures this setting of parameters. Our proof of Theorem \ref{t:efadv-r-inf} does not yield better than $2^n$-size adversaries for $W=\{0,1\}^n$, which is the setting that would correspond to a constructive $\Res(k)$-adversary for $\neg\ttable(f,t)$, so we can match Theorem \ref{t:efadv-r-inf} with Theorem \ref{t:efadv-inf} but not with Theorem \ref{t:efadv-inf2}. 

The proof of Theorem \ref{t:efadv-r-inf} 
requires a detailed analysis of Razborov's original proof 
combined with a different proof of the final width lower bound. We did not manage to exploit \cite{Rkdnf} in a more black-box fashion. Verifying Theorem \ref{t:efadv-r-inf} thus requires reading the entire proof of Razborov (with a more sparse restriction while keeping the original expansion parameters) except the switching lemma which can be treated mostly as a black-box. As a result of this our write-up becomes dense, with many references to the intricacies of the proof in \cite{Rkdnf}. 
 We replace the width lower bound in Razborov's proof by an adaptation of a width lower bound presented by Kraj\'i\v{c}ek \cite[Lemma 13.4.5]{Kpc}. In the adaptation, 
instead of using Kraj\'i\v{c}ek's expansion, we use Razborov's `lossless' expansion.
\medskip

The second example we present is a constructive version of the method of pseudowidth, a sophisticated analogue of the popular method of width \cite{BW}, which was used to derive Resolution (\Res) lower bounds for the weak pigeonhole principle and for formulas encoding circuit (or even DNF) lower bounds \cite{Rwphp,Rpw1,Rpw2,Rpw3,dRpw}. Pseudowidth and width are currently incomparable: It is not known how to prove \Res lower bounds for the weak (enough) pigeonhole principle or for formulas encoding DNF lower bounds using width and it is not known how to prove \Res lower bounds for proof complexity generator formulas $\tau_{W}(A,b)$ considered in Theorem \ref{t:efadv-r-inf} using pseudowidth. 

We present a constructive version of the lower bound for the weak pigeonhole principle $\PHP^{m}_n$ mapping $m$ pigeons to $n$ holes.

\begin{theorem}[Constructive pseudowidth lower bound -- Informal, cf. Theorem \ref{t:epw}]\label{t:efadv-pw-inf}\ \newline 
There is a $poly(m)$-size randomized $\Res$-adversary $C$ claiming that $\neg \PHP^m_n$ with $m=2^{n^{1/5}}$ is satisfiable such that for each $2^{\Omega(n^{1/4})}$-size $\Res$-refutation $\pi$ of any set of clauses, with probability $\ge 9/10$ over the randomness of $C$, 
$\pi$ does not catch $C$.
\end{theorem}

Our proof of Theorem \ref{t:efadv-pw-inf} differs substantially from the proofs in \cite{Rwphp,Rpw1,Rpw2,Rpw3,dRpw} and might be of independent interest. The strategy of the adversary from Theorem \ref{t:efadv-pw-inf} is guided by random partial multi-matchings. The notion of pseudowidth arises here naturally from the observation that the partial multi-matchings satisfy certain clauses with high probability --- these clauses can be thus treated as pseudo-axioms. 

\medskip

Finally, we look at the crown jewel of proof complexity, the lower bound on the length of proofs of the pigeonhole principle $\PHP_n$ (mapping $n+1$ pigeons to $n$ holes) in depth $d$ Frege systems ($\AC_d$-Frege), which is obtained via $k$-evaluations \cite{Aphp,KPW,PBI}. 

Unfortunately, it is not known how to prove a lower bound for $\ttable$-tautologies in $\AC$-Frege. The closest result of this kind was obtained in \cite{Khnw}, where it was shown that a particular kind of $\NP\cap \coNP$ lower bounds (similar to formulas $\neg\tau_W(A,b)$), formalized in a suitable way, does not admit short proofs in \AC-Frege. The lower bound in \cite{Khnw} proceeds by a reduction to the $\PHP_n$ lower bound.

\begin{theorem}[Constructive \AC-Frege lower bound -- Informal, cf. Theorem \ref{t:efadv-keval}]\label{t:efadv-keval-inf}\ \newline 
Let $\epsilon<1/9^d$. There is a $poly(n^{n^{\epsilon}})$-size randomized $\AC_d$-Frege-adversary $C$ claiming that $\neg \PHP_n$ is satisfiable such that for each $2^{n^{\epsilon}}$-size $\AC_d$-Frege-refutation $\pi$ of any set of clauses, with probability $\ge 9/10$ over the randomness of $C$, 
$\pi$ does not catch $C$.
\end{theorem}

The proof of Theorem \ref{t:efadv-keval-inf} follows directly by inspection of the original proof. On high level, the reason why the theorem holds is that a random partial matching `kills' short \AC-Frege proofs of $\PHP_n$ with high probability.

\medskip

\noindent {\it Unification.} Theorems \ref{t:efadv-r-inf}-\ref{t:efadv-keval-inf} imply the corresponding proof complexity lower bounds for $\Res(k)$, \Res and $\AC$-Frege. Moreover, our presentation of lower bound methods of width, pseudowidth and $k$-evaluation in terms of adversaries provides a more unifying perspective on these methods by showing how they arise from natural choices of strategies of adversaries.

\subsection{Related work}\label{s:related}

There are several results in proof complexity that could be interpreted as barriers for deriving proof complexity lower bounds.
\smallskip

\noindent {\em Incompleteness.} Already G\"odel's second incompleteness theorem implies that, e.g., if \ZFC is consistent, then \ZFC cannot prove superlinear lower bounds for the propositional proof system obtained from \ZFC in a natural way (otherwise \ZFC would prove its own soundness). For similar reasons, a result of Kraj\'i\v{c}ek \cite[Corollary 3.2]{Kimp} implies that Buss's theory \SB cannot prove superlinear lower bounds on the so called Implicit \EF. These results, however, do not present an actual barrier for proving proof complexity lower bounds because in proof complexity we prove lower bounds under the assumption that the proof system in question is sound.
\smallskip

\noindent {\em Cook's translation.} A more substantial proof complexity barrier was obtained by Kraj\'i\v{c}ek-Pudl\'ak \cite{KPmod}, who showed that Cook's theory \PV cannot prove that there exists a tautology which requires superpolynomial, say $n^{\log n}$-size, \EF proofs (Buss \cite{Bint} and Cook-Urquhart \cite{CU} derived the same unprovability result for the intuitionistic \PV). The result crucially relies on the formalization of the statement of the lower bound, which is unfortunately not natural from the perspective of \PV in the sense that $n^{\log n}$ is not the length of a number. Consequently, \PV cannot use induction over the proofs of size $n^{\log n}$. Informally, the theory cannot see (and analyze) the whole $n^{\log n}$-size proof properly.
\smallskip

\noindent {\em Nisan-Wigderson generators and the sum-check protocol.} This drawback was eliminated in a proof complexity barrier obtained in \cite{PSapc} (building on \cite{Knw,Pclba} and extended in \cite{LO}), where it was shown that subexponential average-case lower bounds for nondeterministic circuits are not provable in \PV (and certain extensions of \PV). In a related work using completely different methods, Chen, Rothblum and Tell \cite{CRT} constructed a non-deterministic algorithm $P$ such that under a derandomization assumption \SB cannot prove that $P$ is not a p-bounded propositional proof system.\footnote{\cite{CRT} presents the unprovability result for \APC but it works for \SB as well.} Both \cite{PSapc} and \cite{CRT} use a formalization which allows the theory to perform induction over objects of even exponential size $2^{O(n)}$. Unfortunately, the complexity of these formalizations is $\Sigma^p_2$ and while theories like \PV and \SB can prove a lot of complexity theory and are generally very powerful when reasoning with $\Ptime$-time objects and deriving \coNP statements, they are not well-designed for reasoning with \NP statements\footnote{For example, it follows  from Buss's witnessing theorem \cite{Bba} that \SB cannot prove that a {\sf TFNP} problem is total unless the problem is in \Ptime.}, which is why these results do not present a significant barrier in the practice of proving lower bounds.

This issue was partially addressed in \cite{Khjump}. We believe that \cite{Khjump} implicitly shows that there is a concrete {\sf MA} proof system $R$ such that, if it is hard to derive superpolynomial circuit lower bounds in a sufficiently strong propositional proof system $P$, then the theory $T_P$ corresponding to $P$ cannot prove that $R$ does not admit p-size proofs of all tautologies of length $n$. Here, the lower bound for $R$ assumes that $2^{O(n)}$ is the length of a number and it is formulated as a \coNP statement, 
so the $T_P$-provability of the lower bound for $R$ does not necessarily imply that $P$ simulates $R$ and $T_P$ should be in principle very powerful when proving lower bounds for $R$. The result in \cite{Khjump} would therefore give us a remarkable proof complexity barrier if we could only prove the assumption or at least replace it by a standard complexity-theoretic conjecture.

Kraj\'i\v{c}ek \cite{Kht} used the methods of \cite{Knw} to show that under standard hardness assumptions, there is no efficient algorithm which given an algorithm $P$ (potentially defining a propositional proof system) outputs a tautology hard for $P$ or a witness that $P$ does not define a consistent proof system.
\smallskip

\noindent {\em Natural properties.} In an attempt to mimic the natural proofs barrier, \cite{PS19} defined a version of natural properties against proof systems roughly as the efficient provability of \EF lower bounds for formulas $\ttable(f,n^{O(1)})$, for many functions $f$, and  showed that if super-bits exist (an assumption about the existence of pseudorandom generators safe against nondeterministic circuits), then such natural properties against \EF do not exist or \EF does not prove efficiently any formula $\ttable(f,n^{O(1)})$. If we knew how to avoid the second disjunct about the \EF-hardness of $\ttable(f,n^{O(1)})$, this would present a close analogue of the natural proofs barrier in proof complexity.

A combination of the results in \cite{Khjump} and \cite{PS19} shows that under the assumption of the existence of super-bits, either there is no natural proof against $P$ or $P$ cannot prove proof complexity lower bounds for a concrete {\sf MA} system. The conclusion is interesting, but it does not tell us much about the hardness of deriving lower bounds for systems like~\EF.
\smallskip

\noindent {\em Self-provability.} More recently it has been shown that \EF lower bounds imply circuit lower bounds such as $\SAT\notin\Ppoly$ \cite{PSeopt,AKPS} under various assumptions such as that \EF proves efficiently that {\sf E} is hard on average for subexponential-size circuits. If we could avoid this assumption, the resulting strong connection between \EF lower bounds and circuit lower bounds could explain a part of the difficulty of deriving \EF lower bounds.

A problem with interpreting the results from \cite{PS19} or \cite{PSeopt,AKPS} as proof complexity barriers is that they require the assumption that \EF proves efficiently a sufficiently strong circuit lower bound, which is the very statement we are trying to refute when proving proof complexity lower bounds (in Razborov's conjecture). 
\smallskip

\noindent {\em {\sf TFNP}, automatability \& algebraic complexity.} Li, Li and Ren \cite{RLL} initiated a systematic investigation of proof complexity of proof complexity lower bounds through the lens of {\sf TFNP}, but they restrict their attention to Resolution lower bounds, see also \cite{DRlog}. Similarly, the \NP-hardness results for the automatability of various weak proof systems are based on the hardness of deriving proof complexity lower bounds in systems such as Resolution \cite{AM,PAP}. Lu, Santhanam and Tzameret \cite{LST} showed that certain lower bounds for a constant-depth version of \IPS are hard to prove in $\AC[2]$-Frege. The result builds on a work of Santhanam and Tzameret \cite{ST} where \IPS lower bounds for formulas expressing \IPS lower bounds for $\VP\ne \VNP$ were shown to follow from $\VP\ne \VNP$. A drawback of \cite{ST} when interpreted as a barrier result is that the conclusion holds only for infinitely many input lengths, \cite{ST} does not rule out the possibility that \IPS proves efficiently lower bounds for \IPS for $\VP\ne\VNP$ on an infinite and dense set of input lengths. Nevertheless, the potential sensitivity to the input length is a valuable information about the possibility of deriving proof complexity lower bounds for strong proof systems.
\smallskip

\noindent {\em Upper bounds.} Ultimately, the prospect of proving proof complexity lower bounds is limited by the possible existence of surprising upper bounds. In this context, it is worth mentioning an {\sf MA} protocol of Williams which breaks a version of {\sf SETH} for {\sf MA} \cite{Wma}. A recent result of Mazor and Pass \cite{MPass} and  Hirahara, Ilango and Williams \cite{HIW} constructs subexponential-size (deterministic) circuits for more specific problems like \MCSP, but so far all these upper bounds remain in the regime of {\sf SETH} or {\sf ETH}.

\smallskip
All in all, while there have been several previous results that can be seen as some kind of proof complexity barriers, they are significantly less impactful than natural proofs in circuit complexity. In our view, the concept of efficient adversaries introduced in the present paper comes the closest to the potential ``natural proofs barrier" in proof complexity so far despite the fact that the second assumption in Theorem \ref{t:efadv-inf} is similar to the assumption that \EF proves efficiently some circuit lower bound. Theorem \ref{t:efadv-inf2} could turn out as the main barrier, if the existing lower bounds such as the $\Res(k)$ lower bound for $\ttable(f,t)$ can be made more constructive by designing adversaries who are efficient w.r.t. the queries of the prover, not w.r.t. the formula for which the lower bound is proved. 

\subsection{Open problems}\label{s:problems}

\noindent {\bf Constructivity of proof complexity lower bounds.} The main problem we leave open is the question of constructing efficient adversaries for all the remaining major proof complexity lower bounds. As the proofs of Theorems \ref{t:efadv-r-inf}-\ref{t:efadv-pw-inf} show, it is not always obvious how to constructivize an existing lower bound. For example, in Theorem \ref{t:efadv-pw-inf} we only obtained efficient adversaries for the weak pigeonhole principle, but the question of constructing such adversaries for \Res lower bounds for the {\em functional} weak pigeonhole principle \cite{Rpw2,Rpw3} remains open. Further, it is not clear how to get efficient adversaries from lower bounds exploiting `deterministic' restrictions such as the recent super-linear lower bound for constant-depth Frege systems by Gryaznov and Talebanfard \cite{GT}. In general, lower bounds obtained via the width-size relation of Ben-Sasson and Wigderson \cite{BW} use deterministic restrictions as well and we do not know how to make them constructive. That is, we do not have a general strategy for replacing deterministic restricitions whose existence is shown non-constructively by random restrictions which would work with high probability. A prominent example we leave as an open problem is the constructivization of lower bounds based on feasible interpolation such as Pudl\'ak's lower bound for Cutting Planes \CP \cite{Pcp} --- notably, it is still an open problem to show that $\ttable(f,s)$ tautologies are hard for $\CP$. Nevertheless, a significant barrier for using feasible interpolation to derive lower bounds for strong proof systems has been established already in \cite{KP,BPR,BDGMP}.
\begin{problem} Does the \CP lower bound of Pudl\'ak \cite{Pcp} admit efficient adversaries? 
\end{problem}

\noindent If there are lower bounds which cannot be made constructive, can we rule out the possibility of a more substantial proof complexity barrier by showing at least that under a standard complexity-theoretic assumption some proof complexity lower bound does not admit an efficient adversary? 
\vspace{-10pt}

If we cannot construct an efficient adversary for some lower bound, if for each efficient $P$-adversary claiming that $\neg\phi$ is satisfiable, there is a short $P$-refutation (not necessarily of $\neg\phi$) catching the adversary,
we can say that the refutations constitute a short {\em $P$-pseudo-proof} of $\phi$. Is there a proof system $P$ with p-size pseudo-proofs of all tautologies? 
To formulate the question more formally we consider only proof systems $\EF_P$ obtained by extending \EF with axioms postulating the soundness of a proof system $P$. $\EF_P$ simulates $P$ \cite{Kpc}. An $\EF_P$-adversary is an \EF-adversary who is additionaly required to assign 1s to all instantiations of the axiom postulating the soundness of $P$. 

\begin{problem}[$\NP=\coNP$ w.r.t. efficient adversaries] Is there a proof system $P$ such that for each tautology $\phi$, for each $(|\phi|+n)^3$-size $\EF_P$-adversary claiming that $\neg\phi$ is satisfiable there is a $poly(|\phi|)$-size $\EF_P$-refutation catching the adversary
?
\end{problem}

\smallskip

\noindent {\bf Avoiding anticheckers.} Is it possible to remove the assumption about the provably constructive anticheckers from Theorem \ref{t:efadv-inf}? In order to do so, it might be helpful to interpret efficient adversaries as algorithms for the search problems of finding errors in supposed short refutations of hard formulas, cf. \cite{RLL}.
\medskip

\noindent{\bf Efficient adversaries w.r.t. $\ttable$-tautologies and other formulas.} Theorem \ref{t:efadv-inf2} does not rule out the possibility of proving constructively \EF lower bounds for $\ttable(f,s)$ formulas with adversaries which are efficient (not w.r.t. the queries of the prover but) w.r.t. the size of the $\ttable(f,s)$ formula, i.e. with adversaries of size $2^{O(n)}$. Is it possible to make Theorem \ref{t:efadv-inf2} work in this case as well? Can we further extend the theorem to all formulas instead of just formulas expressing circuit lower bounds? The latter question is related to the problem of proving that \MCSP is \NP-complete. Another relevant problem in this direction is to make Theorem \ref{t:efadv-r-inf} work for $\ttable(f,t)$ with better than $2^{O(n)}$-size adversaries, i.e. to construct $\Res(k)$-adversaries for $\neg\ttable(f,t)$ which are efficient w.r.t. the queries of the prover.
\medskip

\noindent{\bf Simulations preserving efficient adversaries.} The notion of efficient adversaries appears to be sensitive to the specific choice of proof system in question. Do p-simulations preserve efficient adversaries? That is, is it possible to show that if there is an efficient $P$-adversary for $\Phi$ and the proof system $P$ is p-equivalent to a proof system $R$, then there is an efficient $R$-adversary for $\Phi$? Is there another natural notion of simulation between proof systems which preserves efficient adversaries?
\medskip

\noindent{\bf Connection to automatability.} 
Automatability of a proof system $P$ should yield a close-to-optimal $P$-adversary because the optimal strategy for the adversary proceeds by deciding in each step if a given query has a short derivation from formulas that have been already assigned 1. Is it possible to use this connection to shed new light on the question of automatability of standard proof systems?

\section{From efficient adversaries to algorithms}\label{s:algfromadv}

In this section we formally introduce efficient adversaries and prove Theorem \ref{t:efadv-inf}.
\bigskip

\noindent {\bf Standard proof systems.} We define adversaries for standard propositional proof systems from the set \SYS consisting of:
\begin{itemize}
\item[$\circ$] Extended Frege \EF, Frege, $\AC_d$-Frege and Circuit Frege \CF systems; 
\item[$\circ$] Resolution \Res; 
Cutting Planes \CP; 
$k$-DNF Resolution $\Res(k)$;
\item[$\circ$] tree-like versions of the systems listed above.\footnote{Strictly speaking, it does not make much sense to consider tree-like \EF as we want to allow the system to reuse extension axioms.}
\end{itemize} 
The definitions of these systems can be found in \cite{Kpc}, but we will use the definition of $\Res(k)$ from \cite[Definition 2.6]{Rkdnf} and the definition of \CF from \cite{Jwphp}. Each instantiation of \EF, Frege, $\AC_d$-Frege and \CF system is represented in \SYS separately. This is because we do not know whether the existence of an efficient adversary for a system $P$ guarantees the existence of an efficient adversary for a system $P'$ if $P'$ is p-equivalent to $P$. We consider Frege, \EF and \CF systems for all finite and complete languages for propositional logic. $\SYS^{-}$ will stand for \SYS without \EF systems.

We denote by $O(P)$ the set of `objects' a proof system $P\in\SYS$ operates with. \EF and Frege systems operate with formulas built from the connectives in the respective language. $\AC_d$-Frege operates with formulas of depth $d$ defined as in \cite[Chapter 15]{Kpc}, \CF systems operate with circuits in the respective language, 
\Res operates with disjunctions of literals (clauses), \CP operates with integer inequalities, $\Res(k)$ with $k$-DNFs. 

In order to unify the presentation, we treat all systems $P\in\SYS$ as refutational systems, refer to the objects of $O(P)$ as formulas (in the case of \CF this is an abuse of terminology but it should not cause any confusion) and denote by 0 the last formula of a refutation.
\medskip

\noindent {\bf Pudl\'ak-Buss game.} Our definition of efficient adversaries is inspired by the Pudl\'ak-Buss game \cite{PB}, which can be used to characterize short Frege proofs. The game is played between Prover and Liar. In each round Prover sends a formula to Liar who responds by 0 or 1. Denote the response of Liar on formula $\psi$ by $C(\psi)$, not displaying the history of the game for simplicity. Intuitively, we interpret $C(\psi)=0/1$ as claims that Liar possesses a falsifying/satisfying assignment of $\psi$. In each round, Liar has to satisfy $C(0)=0$ and $C(\phi)=1$, if we want to estimate the size of the refutation of a formula $\phi$.

Given an $s$-size Frege-refutation of $\phi$, there is a strategy of Prover `catching' Liar in $K\log s$ rounds by querying formulas of size $\le Ks^K$, where $K\in \mathbb{N}$ depends only on the Frege system: Using binary search Prover can find in $K\log s$ rounds formulas $\phi_1,\dots,\phi_{k+1}$ such that $C(\phi_{i})=1$, for $i\in [k]$, $C(\phi_{k+1})=0$ and $\phi_{k+1}$ follows from $\phi_1,\dots,\phi_k$ by a single application of a Frege-rule. 

In the opposite direction, such a strategy of Prover that catches Liar in $r$ rounds with queries of size $\le 2^{O(r)}$ yields a $2^{O(r)}$-size tree-like Frege-refutation of $\phi$, if the logical connectives in the language of the system include $0,1,\neg,\vee,\wedge$, cf. \cite[Lemma 2.2.3]{Kpc}. 

\begin{definition}[Efficient adversary]\label{d:ead}
For $P\in\SYS^-$, an {\em $s$-size $P$-adversary} 
is a sequence $C\colone\{C_i\}_{i=1}^{\infty}$, where $C_i$ computes the responses of the adversary in the $i$-th round as follows. Each $C_i$ is a sequence of $s$-size circuits $C_i\colone\{C_{i,j}\}$ which for each $\left<a_1,\dots,a_{i}\right>\in\mathbb{N}^i$ contains a single circuit $C_{i,j}$ with $i$ blocks of inputs partitioned so that the $k$-th block, $k\le i-1$, consists of $a_k+1$ inputs and the $i$-th block consists of $a_i$ inputs. Given a history of the adversary's responses $h\colone\{\left<\phi_k,b_k\right>\}^{i-1}_{k=1}$, where $\phi_k\in O(P)$ and $b_k\in\{0,1\}$ is a response of the adversary to $\phi_k$ in the $k$-th round, the response of the adversary to $\phi_i\in O(P)$ in the $i$-th round is $C_i(h,\phi_i)\colone C_{i,j}(h,\phi_i)$, where $C_{i,j}$ is the circuit in $C_i$ with inputs partitioned in accord with $h,\phi_i$. 
The $P$-adversary is {\em honest} if it responds consistently with its responses in previous rounds, i.e. for each $\phi\in O(P)$ and each history $h=\{\left<\phi_k,b_k\right>\}^{i-1}_{k=1}$, $C_i(h,\phi)=b_{\ell}$ if $\phi=\phi_{\ell}$ for some $\left<\phi_{\ell},b_{\ell}\right>\in h$ with $\ell<i$. 

If the adversary responds to $\phi\in O(P)$ with a fixed $b\in\{0,1\}$ in each round independently of the history, we write $C(\phi)=b$. 
If the $P$-adversary satisfies $C(0)=0$ and $C(\phi)=1$, for all $\phi\in\Phi\subseteq O(P)$, we say that the adversary is {\em for} $\Phi$. We will consider only finite sets $\Phi$.
\medskip

A {\em randomized $s$-size $P$-adversary} ({\em for} $\Phi$) is an $s$-size $P$-adversary $\{C_i\}_{i=1}^{\infty}$ ({\em for} $\Phi$) such that for each $i$, each $C_{i,j}\in C_i$ has access to $\le s$ bits of a finite random string of bits shared among all circuits $C_{i,j}$, for all $i,j$. The random strings are distributed according to some arbitrary distribution. 
\end{definition}

The definition of \EF-adversaries requires a small twist which will become apparent after Definition \ref{d:evalead}. For $P\in\SYS^-$, a $P$-prover is a function mapping histories $\{\left<\phi_k,b_k\right>\}_{k=1}^i$, where $\phi_k\in O(P)$, $b_k\in\{0,1\}$, $i\in\mathbb{N}$, to new formulas (queries) in $O(P)$.
\medskip

\noindent{\bf Switching quantifiers and the role of randomness.} If for a given prover with $\le s$-size queries, there is an adversary who stays consistent for $r$ rounds, then there is a $poly(sr)$-size adversary who achieves that as well. To make the notion of efficient adversaries interesting we will thus switch quantifiers and consider $s$-size adversaries staying consistent for $r$ rounds against all possible provers. A randomized adversary will be slightly more flexible: for each prover, the randomized adversary will choose a random string and then proceed as the deterministic adversary specified by the random string. 
The definition ensures that circuits in $C_2$ responding in the second round can access the random bits used by circuits $C_1$ responding in the first round.
\smallskip

Next we introduce a notion that allows us to generalize the characterization of proof-size from Pudl\'ak-Buss games to all systems in $\SYS^-$.

\begin{definition}[Evaluation of adversary]\label{d:evalead} Given a refutation $\pi$ of $\Phi\subseteq O(P)$ in a proof system $P\in\SYS^-$ and an ($s$-size) $P$-adversary $C$ for some $\Phi'\subseteq O(P)$, define the {\em evaluation of $C$ on $\pi$} by the following process. In round 0, let $h_0\colone\emptyset$ and start at the last formula of $\pi$, i.e. 0. If in the $i$-th round, the current formula $\phi\notin\Phi$ is derived in $\pi$ by a single application of a derivation rule from (a possibly empty set of) formulas $\phi_j, j\in [k]$, proceed to the first $\phi_{j'}, j'\in [k]$, such that $C_{i+1}(h_i,\phi_{j'})=0$ and let $h_{i+1}\colone h_i\cup \left<\phi_{j'},0\right>$. If there is no such $\phi_{j'},j'\in [k]$, output 0. Similarly, if the current formula $\phi\in\Phi$ is derived in $\pi$ by a single application of a derivation rule from a non-empty set of formulas $\phi_j,j\in [k]$, proceed to the first $\phi_{j'}, j'\in [k[$ such that $C_{i+1}(h_i,\phi_{j'})=0$ and output 0 if there is no such $\phi_{j'}, j'\in [k]$. 
If an initial formula $\phi\in\Phi$ in $\pi$ is reached, output 1. If $\pi$ evaluated $C$ to 0, we say that $\pi$ {\em caught} $C$. Otherwise, $C$ {\em stayed consistent on} $\pi$.

\end{definition}

Obviously, there is an $s$-size refutation of $\Phi$ in a proof system $P\in\SYS^-$ if and only if there is an $s$-size 
$P$-refutation of $\Phi$ 
which catches all 
$P$-adversaries for $\Phi$. 

In general, an adversary evaluated on a proof has to respond to a premise of an application of a derivation rule without knowing the remaining premises. If we gave the adversary access to all premises, Theorems \ref{t:efadv-inf}-\ref{t:efadv-inf2} would remain valid. It turns out that efficient adversaries constructed in practice (Theorems \ref{t:efadv-r-inf}-\ref{t:efadv-keval-inf}), do not need to receive all premises.
\medskip

\noindent {\bf \EF adversaries.} Definitions \ref{d:ead}-\ref{d:evalead} can be straightforwardly adapted to the case of \EF systems, but the extension axioms need to be treated carefully: It would be unfair to ask an \EF adversary to assign a value to an extension variable $e$ and only subsequently reveal that $e$ is equivalent to 1 or 0 --- the adversary would be caught immediately. The issue does not appear in the case of the p-equivalent \CF systems. This also suggests the right way to define \EF adversaries: We adapt Definitions \ref{d:ead}-\ref{d:evalead} so that whenever an \EF adversary evaluated on an \EF refutation $\pi$ receives a formula $\phi$, they receive also all extension axioms in $\pi$ defining the extension variables appearing in $\phi$ and the extension axioms of all the extension variables in the received extension axioms. In other words, the adversary gets to know the whole circuit defining the extension variables in $\phi$.

To justify the definition, following the proof of the simulation of \EF by \CF \cite[Lemma 2.4]{Jwphp}, it can be shown that for each \EF system $P$ and each \CF system $R$ such that $P$ and $R$ have the same language and Frege rules, if each $poly(s)$-size $P$-adversary is caught by some $t$-size $P$-refutation, then each $s$-size $R$-adversary is caught by some $poly(t)$-size $R$-refutation.
\medskip

Proposition \ref{p:fadv} points out that for Frege systems, an honest adversary for $\Phi$ stays consistent against all provers with $poly(n)$-size queries for $O(\log n)$ rounds if and only if it stays consistent on all $poly(n)$-size $O(\log n)$-height refutations (not necessarily of $\Phi$). (Note that 
an $s$-size Frege proof can be transformed to a $poly(s)$-size $O(\log s)$-height Frege proof of the same formula \cite{Kpc}.) 
A height of a Frege derivation is the maximal number of edges on a directed path in the derivation graph.


\begin{proposition}\label{p:fadv} Let $P$ be a Frege system containing modus ponens rule (which derives $\psi$ from $\phi$ and $\neg\phi\vee\psi$). 
\begin{itemize}
\item[1.] If an $s$-size $r$-height $P$-refutation (not necessarily of $\Phi$) catches a $P$-adversary for $\Phi$, then some $P$-prover with $s$-size queries catches the adversary in $O(r)$ rounds. 

\item[2.] If a $P$-prover with $s$-size queries catches an honest $P$-adversary for $\Phi$ in $r$ rounds, then some $poly(s,r)$-size $(r+2)$-height $P$-refutation (not necessarily of $\Phi$) catches the adversary.
\end{itemize}
\end{proposition}

\proof To prove Item 1, we define a strategy of the prover by querying formulas on the path in the $P$-refutation going from 0 to an axiom where the adversary assigns 0s. 

Item 2 is obtained by treating each query as a resolvent $\phi$ in the modus ponens rule deriving $\psi$ from $\phi$ and $\neg \phi\vee \psi$. 
The formula $\psi$ is initially 0. If $\phi$ is assigned 0, we proceed with extending the refutation we have built so far by building a derivation of $\phi$ and leaving $\neg \phi\vee \psi$ as an additional axiom. Otherwise, we proceed with building a derivation of $\neg\phi\vee\psi$ and leaving $\phi$ is an axiom. The evaluation of the adversary on the resulting refutation $\pi$ ends up in one of the following two cases.  In Case 1, the last queried formula $\phi_r$ is assigned 0 but $\phi_r$ follows by a single application of a Frege rule from formulas assigned 1. In Case 2, the last queried formula is assigned 1 and again a `direct contradiction' is found in which some formula $\phi_r$ is assigned 0. In Case 1, we extend $\pi$ by a single derivation step deriving $\phi_r$ from the formulas assigned 1. In Case 2, we extend $\pi$ by a modus 
 ponens rule resolving on $\phi_r$ and then adding the derivation rule deriving $\phi_r$ from the formulas assigned 1. Since the adversary has to stay consistent with its previous responses, in both cases, we obtain an $\le (r+2)$-height $P$-refutation of size $poly(s,r)$ which catches them. \qed

\bigskip



Proposition \ref{p:fadv} holds also for $\AC_d$-Frege systems with a minor adjustment of parameters. 
We do not know how to improve the conclusion in Proposition \ref{p:fadv} (Item 2) into a $P$-refutation of $\Phi$ catching the adversary, 
but we will show that a typical proof complexity lower bound for a formula $\Phi$ yields a randomized efficient (possibly dishonest) adversary for $\Phi$ who stays consistent on all small refutations (not necessarily of $\Phi$) with probability $\ge 9/10$ over the randomness of the adversary, if the adversary is not required to assign 1s to formulas outside $\Phi$. This should be contrasted with the next theorem showing that under certain assumptions it is not possible to construct such \EF-adversaries for formulas encoding circuit lower bounds for \SAT.

In order to state the theorem we need to introduce some notation. Let $\neg\lb_{A}(\SAT,s)[B]$ be a propositional formula encoding the $poly(s,|A|)$-size predicate $R_1(B)\hspace{2pt}\colone$ ``variables $B$ encode an $s$-size circuit with $n$ inputs computing \SAT on inputs from $A\subseteq\{0,1\}^n$". The formula can use other hidden auxiliary variables as well and is satisfiable iff $\exists B, R_1(B)=1$. The values of $\SAT$ on $A$ are `hardwired' bits. Sometimes we write $\lb_A(\SAT,s)$ without displaying $[B]$. We define $\ttable(\SAT,s)$ as $\lb_{\{0,1\}^n}(\SAT,s)$. The notation \SAT should not be confused with $\SAT(x)$ which we use as a shortcut for saying that ``$x$ is a satisfiable formula" and with $\SAT(x,y)$ used to denote a propositional formula encoding the predicate $R_2(x,y)\hspace{2pt}\colone$ ``$x$ encodes a propositional formula satisfied by assignment $y$". For concreteness, we assume that $\SAT(x,y)$ uses extension variables encoding the gates of a $poly(|x|,|y|)$-size circuit computing $R_2(x,y)$ and is a tautology for an assignment $a,b$ of $x,y$ iff $R_2(a,b)=1$. Similarly, given a circuit $F$ with inputs $y$, the predicate $R_3(x,y)\hspace{2pt}\colone$ ``$x$ is a propositional formula satisfied by $F(y)$" can be encoded by a propositional formula $\SAT(x,F(y))$ using extension variables for the gates of a $poly(|F|,|x|)$-size circuit computing $R_3(x,y)$ and is a tautology for an assignment $a,b$ of $x,y$ iff $R_3(a,b)=1$. We do not specify the encoding of these formulas further as the results in this section work for any natural encoding. 



\begin{theorem}\label{t:efadv}\ \newline 
For each \EF system $P$ there is $K\in\mathbb{N}$ such that the following holds.

Let 
$v:\mathbb{N}\mapsto\mathbb{N}$, $n,s,t,u,w\in\mathbb{N}$ be such that $n<s\le t\le u$ and assume 1 - 2: 
\begin{itemize}
\item[1.] {\em (The canonical pair of $P$ is hard.)} For each $(n\cdot v(t^K))^K$-size circuit $D$ with $n$ inputs, there is $z$ such that $\SAT(z)\wedge D(z)=0$ or $D(z)=1$ while there is a $w$-size $P$-refutation of (the formula encoded by) $z$. 
\item[2.] {\em (Provably constructive anticheckers.)} There is a $t$-size set $A\subseteq\{0,1\}^n$ and a $t$-size circuit $F$ such that for each $z\in\{0,1\}^n$ there is a $u$-size $P$-proof of $$\neg\lb_A(\SAT,s)[B]\wedge\SAT(z,y)\rightarrow \SAT(z,F(z,B)).$$
\end{itemize}
Then,

\begin{itemize}

\item[a)] For each (possibly dishonest) randomized $v$-size $P$-adversary $C=\{C_i\}_{i=1}^{\infty}$
for $\neg\lb_A(\SAT,s)$, there is a $(w+u)^{K}$-size $P$-refutation catching $C$ with probability $\ge 1/2-1/n$ 
over the randomness of $C$. 

\item[b)] 
For each randomized $v$-size $P$-adversary $C$ for $\neg\ttable(\SAT,s)$, there is a $(2^n+w+u)^{K}$-size $P$-refutation catching $C$ with probability $\ge 1/2-1/n$ over the randomness of~$C$. 
\end{itemize}

Moreover, Item $b)$ holds even without the assumption in Item 2, if 
$s\le 2^n$.

\end{theorem}


Typical parameters of interest in Theorem \ref{t:efadv} are $s,t,u=poly(n)$ and $v$ being a polynomial function in $n$ and the size of the input, where the size of the input might be $\ll n$ or $\gg n$. In Item $a)$, we typically consider $w=poly(n)$ while in Item $b)$ it is best to set $w=2^{O(n)}$.

\begin{remark}\label{r:moreover} For $w=2^{O(n)}$, which is the setting suitable for the `moreover' part of Theorem~\ref{t:efadv}, Item 1 in the assumptions can be replaced by $\SAT\notin\Circuit[(n\cdot v(t^K))^K]$ because each unsatisfiable $z$ has a $2^{O(|z|)}$-size \EF-refutation.
\end{remark}

\proof[Proof of Theorem \ref{t:efadv}]\ \newline

\noindent Item $a)$. 
Assume that the modus ponens rule is available in $P$ and let $D^r$ be the $n$-input $v(poly(t))$-size randomized circuit $$D^r(x)\colone\hspace{1pt} C_{1,j}(\SAT(x,F(x,B))),$$ where $C_{1,j}$ is the randomized circuit in $C_1$ with the appropriate input size, $F$ is the circuit from the second assumption and $B$ are the variables from the displayed formula in the second assumption. In more detail, $D^r$ maps $x\in\{0,1\}^n$ to $\SAT(x,F(x,B))$ and sends the formula to $C_{1,j}$ together with the extension axioms of the extension variables in $\SAT(x,F(x,B))$. Intuitively, $D^r$ attempts to use $C$ to solve \SAT. However, there is an input $z$ of $D^r$ on which $D^r$ fails to separate the canonical pair of $P$ (for the given $w$) with probability $\ge 1/2-1/n$. Otherwise, by an amplification employing Chernoff's bound and the subsequent derandomization, a circuit of size $poly(n)\cdot |D^r|+poly(n)$ would separate the canonical pair of $P$, contradicting Item~1. 

The last step of the refutation catching $C$ derives 0 from $\SAT(z,F(z,B))$ and its negation (using modus ponens).\footnote{If we formulated the proof in terms of prover-adversary games, $\SAT(z,F(z,B))$ would be the first formula sent by the prover.} By the choice of $z$, with probability $\ge 1/2-1/n$ over the randomness of $C$, one of the following two cases happens.
\medskip

\noindent Case I: $D^r(z)=1$ and there is a $w$-size $P$-refutation of $z$. A $w$-size $P$-refutation of $z$ yields a $poly(w+t)$-size $P$-derivation $\pi_0$ of $\neg\SAT(z,F(z,B))$ from empty assumptions, cf. \cite[Chapter 12.4]{Kpc}. Since extension variables cannot appear in the initial formulas of a $P$-refutation containing $\pi_0$ (they would not be allowed to be introduced in $\pi_0$), we also need to use a $poly(t)$-size $P$-derivation $\pi_2$ of $\SAT(z,F(z,B))$ from $0$. The refutation catching $C$ is then constructed by composing $\pi_0,\pi_2$ with the application of modus ponens on $\SAT(z,F(z,B))$ and its negation. It is a refutation of $0$. 
\medskip

\noindent Case II: $D^r(z)=0\wedge\SAT(z)$. Since $\SAT(z)$ and we assume Item 2, there is a $poly(u)$-size $P$-derivation $\pi_1$ of $\SAT(z,F(z,B))$ from $\neg\lb_A(\SAT,s)$. Again, in order to deal with the extension axioms, we construct a $poly(t)$-size $P$-derivation $\pi_2'$ of $\neg\SAT(z,F(z,B))$ from $0$ and then the $P$-refutation catching $C$ analogously as in Case I. It is a refutation of $\neg\lb_A(\SAT,s)$ and $0$. 
\medskip

\noindent Since only one of the cases I and II can happen, this yields the desired refutation. 

If modus ponens is not available in $P$, we proceed similarly. Given $x\in\{0,1\}^n$, $D^r$ constructs a $poly(t)$-size $P$-derivation $\pi_x$ of 0 from $\SAT(x,F(x,B))$ and its negation (expressed in the language of the system). Then, $D^r$ simulates the evaluation of $C$ on $\pi_x$ (it suffices to store finitely many $v(poly(t))$-size circuits from $C$ to simulate the evaluation for all $x\in\{0,1\}^n$). If the evaluation ends up in the formula $\SAT(x,F(x,B))$, $D^r$ outputs 0. If the evaluation ends up in the formula $\neg\SAT(x,F(x,B))$, $D^r$ outputs 1. Otherwise, $D^r$ outputs an arbitrary bit, say, 0. The size of $D^r(x)$ is $poly(v(poly(t)))$. The existence of $z\in \{0,1\}^n$ such that Case I or Case II happens follows as before. The refutation catching $C$ is constructed as before but the final application of modus ponens, i.e. the last step of the refutation, is replaced by $\pi_z$. The refutation catches $C$ because in Case I, the evaluation of $C$ on $\pi_z$ cannot end up in $\SAT(z,F(z,B))$ --- in such case $D^r$ would output 0 on $z$ --- and in Case II the evaluation on $\pi_z$ cannot end up in $\neg\SAT(z,F(z,B))$ --- in such case $D^r$ would output 1 on $z$. This completes the proof of Item $a)$.
\bigskip


\noindent Item $b)$. 
We proceed as in Item $a)$ but instead of the assumption from Item 2 we use a formalization of the search-to-decision reduction for \SAT. In more detail, let $F$ be a $poly(s)$-size circuit which solves the search version of \SAT given an $s$-size circuit $B$ witnessing that $\neg\ttable(\SAT,s)$. As before, the last step of the refutation resolves on $\SAT(z,F(z,B))$, for a suitable $z$ (or this step is simulated). The only difference from Item $a)$ is that in Case II we use a $2^{O(n)}$-size $P$-derivation of $\SAT(z,F(z,B))$ from $\neg \ttable(\SAT,s)$. The existence of the $P$-derivation follows from the \SB-formalization of the search-to-decision reduction for \SAT, see \cite[Proposition 3.1]{Pclba} or \cite{CK}, and the fact that $\Pi^b_1$-consequences of \SB have p-size proofs in \EF \cite{Kba}. 
 \qed

\bigskip

While the assumption in Item 1 in Theorem \ref{t:efadv} is standard (roughly corresponding to the non-automatability of $P$), Item 2 deserves a discussion.
\medskip

\noindent {\bf Plausibility of provably constructive anticheckers.} A theorem of Lipton and Young \cite{LY} about `anti-checkers' guarantees that if \SAT is hard for circuits of size $poly(s)$, there is a set $A$ of size $poly(s)$ such that $\lb_A(\SAT,s)$ is a tautology. In fact, a simple case analysis (considering the cases that $\SAT\in\Circuit[poly(s)]$ and $\SAT\notin\Circuit[poly(s)]$) yields a circuit $F$ with $t=poly(s)$ such that the displayed formulas in Item 2 are tautologies, but we do not have the guarantee that the tautologies are efficiently provable. The existence of such $F$ with the efficiently provable property from Item 2 follows, if we assume that \EF proves efficiently that a concrete Boolean function is hard on average and that \EF proves efficiently that a concrete efficient circuit given a circuit breaking a one-way permutation (OWP) produces a circuit for \SAT, see \cite[Lemma 2]{PSeopt}. In other words, if \EF can prove a circuit lower bound and reduce OWPs to $\SAT\notin \Ppoly$, Item 2 holds. In fact, it holds with $z$ being a free variable in the displayed formula. 

Item 2 is similar to the assumption used in \cite{PSeopt} to establish the implication that lower bounds on lengths of proofs in concrete propositional proof systems imply explicit circuit lower bounds. Indeed, as pointed out to us by Noel Arteche, Item 2 (with $s,t,u=poly(n)$) implies that $\NP\not\subseteq\Circuit[n^{\ell}]$, for each $\ell\in \mathbb{N}$ --- $\SAT\in\Ppoly$ combined with Item 2 implies that \EF is p-bounded and $\NP=\coNP$ implies the desired lower bound using Kannan's lower bound. It is possible that Item 2 implies that \EF proves efficiently $\ttable(f,n^{O(1)})$ for some $f$, but it is not immediate: The outlined proof of $\NP\not\subseteq\Circuit[n^{\ell}]$ involves a case analysis yielding $\ttable(\SAT_{n^k}, n^{\ell})$ or $\ttable(f',n^k)$ where $f'$ has inputs of length $n$. Such disjunctions have, however, trivial proofs of polynomial size in the size of the formula (i.e. in $2^{O(n^k)}$). Nevertheless, Item 2 implies $poly(u)$-size \EF-proofs of $\lb_A(\SAT,s)\vee \ttable(h,s')$, for each $h\notin\Circuit[s']$, where the truth-table of $h$ has length $<n$. Further, by truncating the inputs from $A$ we get $poly(u)$-size \EF-proofs of $\ttable(h',s)\vee \ttable(h,s')$, for some $h'$ with the truth-table of size $\sim t$.\footnote{If we assume the existence of efficient \EF-proofs of $\ttable(h',s)\vee \ttable(h,s')$, for some $h',h$ and suitable $t,s,s'$, then a version of \EF-natural proofs from \cite{PS19} (defined as the existence of efficient \EF-proofs of the formulas encoding that ``\EF does not prove efficiently $\ttable(h',s)\vee\ttable(f,s'/3)$", for many functions $f$) breaks superbits. In order to see this, note that short \EF-proofs of $\ttable(h',s)\vee\ttable(h,s')$ yield short \EF-proofs of $\ttable(h',s)\vee\ttable(h\oplus f,s'/3)$, for every $f\in\Circuit[s'/3]$.}

Importantly, our theorem holds for stronger systems than \EF as well --- specifically, for systems $\EF_P$ from Section \ref{s:problems} --- in case these are needed to satisfy Item 2. One way to interpret Theorem \ref{t:efadv} is thus to say that if we can construct efficient adversaries for lower bounds in a sufficiently strong proof system which can prove constructively the existence of anti-checkers (e.g. by proving some circuit lower bound and reducing OWPs to $\SAT\notin\Ppoly$), then we can separate the canonical pair of the system   efficiently.

\begin{remark}\label{r:phi} Theorem \ref{t:efadv} remains valid if the formula $\neg\lb_A(\SAT,s)$ in its conclusion is replaced by an arbitrary formula $\phi$ such that $P$ has a $poly(u)$-size derivation of $\neg\lb_A(\SAT,s)|_{\rho}$ from $\phi$, where $\psi|_{\rho}$ is a formula obtained from $\psi$ by applying a substitution $\rho$ which replaces some (possibly all) variables of $\psi$ by $poly(s)$-size formulas built on variables of $\phi$. In this case, 
the $P$-refutation is constructed as in the proof of Theorem \ref{t:efadv} but applying $\rho$ on formulas $\lb_A(\SAT,s)$, $\SAT(x,F(x,B))$ and using the fact that the assumption from Item 2 remains valid under $\rho$. Moreover, in Case II, we use the short $P$-derivation of $\neg\lb_A(\SAT,s)|_{\rho}$ from $\phi$. 
Similarly, $\neg\ttable(\SAT,s)$ in Theorem~\ref{t:efadv} can be replaced by an arbitrary formula $\phi$ such that $P$ has a $2^{O(n)}$-size derivation of $\neg\ttable(\SAT,s)|_{\rho}$ from $\phi$, where $\rho$ substitutes some variables by $poly(s)$-size (not $2^{O(n)}$-size) formulas built on variables of $\phi$. 

As noted by Moritz M\"uller, Theorem \ref{t:efadv} holds also if we substitute $\neg\lb_A(\SAT,s)$ both in the conclusion and in the assumption by an arbitrary formula.
\end{remark}

\section{Efficient adversaries for known lower bounds}

\subsection{Width and expansion}\label{s:width}

The strongest known lower bound on the length of proofs of $\ttable$-tautologies is due to Razborov \cite[Theorem 2.13]{Rkdnf}, who showed that (a concrete natural encoding of) $\ttable(f_n,t)$, $n^2\le t\le 2^n$, and in fact also $\lb_W(f_n,t)$, for any $W\subseteq\{0,1\}^n$ and any $f_n$, requires $\Res(k)$-proofs of size $2^{t^{\Omega(1)}}$, if $k=\epsilon\log t$ and $\epsilon>0$ is a sufficiently small constant. The lower bound combines a small-restriction switching lemma, which turns short proofs into narrow ones, with a width lower bound which is obtained using expansion properties of the \lb-formula (similarly as in the influential work of 
Ben-Sasson and Wigderson \cite{BW}). 
We show how to adapt the proof so that it yields efficient adversaries for formulas $\tau_{W}(A,b)$ described below. On the way, we will refer to the details of the proof (and definitions) from \cite{Rkdnf}. 

The reason why we focus on formulas $\tau$ instead of $\lb$ is that Razborov's lower bound for $\lb_W(f_n,t)$ is obtained by showing that every $S$-size $\Res(k)$-refutation of $\neg\lb_W(f_n,t)$ can be turned into an $S2^{O(k)}$-size $\Res(2k)$-refutation of $\tau_{W}(A,b)$ and then proving that $\tau_{W}(A,b)$ does not have such refutations. In fact, as we will observe, $\Res(2)$ admits $poly(|W|,t)$-size derivations of each clause of $\neg\lb_W(f_n,t)|_{\rho}$ from $\tau_{W}(A,b)$, where $\rho$ substitutes some variables of $\neg\lb_W(f_n,t)$ by a function of at most two variables of $\tau_{W}(A,b)$. On the other hand, by Remark \ref{r:phi}, Theorem \ref{t:efadv} holds even if $\neg\lb_W(\SAT,s)$ is replaced by an arbitrary formula $\phi$ such that \EF has a $poly(s)$-size derivation of $\neg\lb_W(\SAT,s)|_{\rho}$ from $\phi$.\footnote{If $\neg\phi$ is to be hard for a strong proof system, it should be different from $\neg\tau_{W}(A,b)$, because systems simulating Gaussian elimination can prove $\neg\tau_W(A,b)$ efficiently.} The focus on $\tau$-formulas allows us to avoid the problem of encoding the transformation of refutations of $\neg\lb$ to refutations of $\tau$ in the strategy of the adversary.
\bigskip

In the following paragraphs defining formulas $\tau_W(A,b)$ we refer mainly to \cite[Sections 6 and 8]{Rkdnf}, where the reduction of lower bounds for $\neg\lb_W(f_n,t)|_{\rho}$ to lower bounds for $\tau_W(A,b)$ is described for the case $W\colone\{0,1\}^n$. We start with specifying formulas $$\tau_{\le}(A,b)=\hspace{-3pt}\vcenter{\hbox{:}}\ \tau_{\{0,1\}^n}(A,b)$$ from \cite{Rkdnf}, for $A,b$ such that the desired short derivation of $\neg\ttable(f_n,t)|_{\rho}$ from $\tau_{\le}(A,b)$ exists. Formulas $\tau_{\le}(A,b)$ are defined generally in \cite[Definition 2.2]{Rkdnf} but we will consider only their particular instantiations obtained from $\tau(C^{n+1}_{A,\le}, b)$ in \cite[Section 8]{Rkdnf} with two caveats: 
\begin{itemize}
\item[1.] We adopt the convention from the beginning of \cite[Section 5]{Rkdnf} introducing axioms $\neg y^i_{\emptyset}$ and $y^i_{\{j\}}\vee \neg x_j, \neg y^i_{\{j\}}\vee x_j$ (instead of identifying $x_j$ with $y^i_{\{j\}}$); 

\item[2.] $A,b$ in our $\tau_{\le}(A,b)$ is not the same as $A,b$ in $\tau(C^{n+1}_{A,\le}, b)$ from \cite[Section 8]{Rkdnf}. In order to specify our $A,b$ in $\tau_{\le}(A,b)$, we need to express $\tau(C^{n+1}_{A,\le},b)$ from \cite[Section 8]{Rkdnf} as in the proof of \cite[Theorem 2.12]{Rkdnf} which in turn refers to the proof of \cite[Theorem 2.10]{Rkdnf}, where $\tau(C^{n+1}_{A,\le},b)$ is interpreted as a linear system $\hat{A} X=\hat{b}$ with constant right-hand side --- this $\hat{A},\hat{b}$ is our $A,b$ in $\tau_{\le}(A,b)$ and our formula $\tau_{\le}(A,b)$ is the formula $\tau_{\hat{\le}}(\hat{A},\hat{b})$ in the proof of \cite[Theorem 2.10]{Rkdnf} obtained from $\tau(C^{n+1}_{A,\le},b)$ considered in \cite[Section 8]{Rkdnf}.
\end{itemize}

The formulas $\neg\tau_{\le}(A,b)$ express that a bit-string $b$ of  the length $|b|=2t_0(2^{n+1}-1)$, $t_0=\Omega((t/n)^{1/2})$, is not in the range of an NW-generator based on a $|b|\times t_0(2^{n+1}-1)$ matrix $A$ and the parity function. In other words, $\tau_{\le}(A,b)$ is a set of clauses asserting that $A\overline{x}=b$ is a system of linear equations solved by a $(|b|/2)$-bit vector $\overline{x}$. The matrix $A$ is an $(r,d+2)$-lossless expander, with $r=t^{\delta}$ for an arbitrary sufficiently small $\delta>0$ and a sufficiently big $d=O(1)$, corresponding to an iterated composition of a function $H:\{0,1\}^{t_0}\mapsto\{0,1\}^{2t_0}$ (iteratively applied on the first $t_0$ and the last $t_0$ of its output bits) as described in \cite[Section~6]{Rkdnf} ($H$ denotes a different object in \cite{Rkdnf}). Additionally, $A$ (is constructed from an $(r,d)$-lossless expander in \cite[Section 8]{Rkdnf} and so it) satisfies $|J_i(A)|\ge t^{\delta}$ for all $i\in [|b|]$.\footnote{The existence of such $A$ follows from the proof of \cite[Theorem 2.5]{Rkdnf} and an application of Chernoff's bound, if $t\ge n_0$ for some $n_0$ depending on $\delta$ and $d$.} The iterated composition of functions $H$ forms a tree $T_H$ of $2^{n+1}-1$ copies of $H$. The indices of $b$ are in one-to-one correspondence with outputs of the copies of $H$ in $T_H$. The string $b$ has 0s on all positions corresponding to outputs of the `inner' copies of $H$. In order to get a short derivation of $\neg\ttable(f_n,t)|_{\rho}$ from $\tau_{\le}(A,b)$ one defines $b_{a,j}\colone f_n(a)$, for $a\in\{0,1\}^n, j\in[2t_0]$, where $b_{a,j}$ corresponds to the $j$th output of the $a$th `outer' copy of $H$, i.e. the copy of $H$ forming the $a$th leaf in $T_H$. As explained in \cite[Section 8]{Rkdnf}, there is a $t$-size circuit $D_{n,x_1,\dots,x_{t_0}}(a)$ computing $b_{a,1}=f_n(a)$ on $a\in \{0,1\}^n$. The circuit $D_{n,x_1,\dots,x_{t_0}}(a)$ gradually computes all copies of $H$ in $T_H$ that lead to $b_{a,1}$. In particular, the $a$th outer copy of $H$ is computed by $D_{n,x_1,\dots,x_{t_0}}(a)$ as well.

Given $W\subseteq \{0,1\}^n$, the formula $\tau_W(A,b)$ is obtained from restricting $\tau_{\le}(A,b)$ to (the axioms encoding) the equations from $A\overline{x}=b$ which are used by the computation of $D_{n,x_1,\dots,x_0}(a)$, for $a\in W$. Consequently, the size of $\tau_{W}(A,b)$ is $poly(t,|W|)$.

By inspecting the reduction in \cite[Section 8]{Rkdnf}, it can be verified that each clause of formula $\neg\lb_W(f_n,t)|_{\rho}$ has a $\Res(2)$ derivation of size $O(1)$ from $\tau_W(A,b)$, where $\rho$ substitutes some variables of $\neg\lb_W(f_n,t)$ by a function of at most two variables of $\tau_W(A,b)$ and $b$ is defined from $f_n$ as above. 
\medskip

In the rest of this section we give a proof of the following theorem in which we construct a $poly(t^{t^{\delta}},m')$-size (randomized circuits defining the responses of an) adversary for $\tau_{W}(A,b)$. 

\begin{theorem}[Efficient adversary for Razborov's $\Res(k)$ lower bound]\label{t:efadv-r}\ \newline 
Let $\gamma<1$, $a\in \mathbb{N}\setminus \{0\}$, let $\delta>0$ be sufficiently small, $d\in \mathbb{N}$ be sufficiently big and $\epsilon>0$. There is $n_0$ such that for all $n\ge n_0$ the following holds. Suppose $A,b,t$ are as above (i.e. $n^2\le t\le 2^n$, $f_n$ specifying $b$ is arbitrary, $A$ is an $(r,d+2)$-lossless expander with $r=t^{\delta}$ and dimensions $|b|\times t_0(2^{n+1}-1)$), $k=\epsilon \log t$ and $W\subseteq \{0,1\}^n$ has size $|W|\le t^a$. There is a $poly(t^{rad},m')$-size randomized $\Res(k)$-adversary $C$ for $\tau_{W}(A,b)$ such that for each $\frac{1-\gamma}{8}2^{r/(2^{O(k(d+2))})}$-size $\Res(k)$-refutation $\pi$ of any set of clauses $\Phi$
, with probability $\ge \gamma$ over the randomness of $C$, $\pi$ does not catch $C$. Here, $m'$ refers to the input length of circuits from $C$.
\end{theorem}

Let $A^-\overline{x}=b^-$ be the system of $m\colone|b^-|$ equations encoded by $\tau_{W}(A,b)$. 
That is, $A^-$ results from $A$ by erasing some rows and $b^-$ is the corresponding restriction of $b$.
Since $m\le |\tau_W(A,b)|=poly(t,|W|)$ and $|W|\le t^a$, such $A^-$ satisfies \begin{equation}\label{e:assumption}|J_i(A^-)|\ge t^{\delta}\ge C'_{\gamma}(d+2)(k+\log m),\text{ for all }i\in [m],\end{equation} for arbitrary constant $C'_{\gamma}$, if 
 $n$ is sufficiently big. 
\medskip

On high level, the strategy of our adversary is to transform $\pi$ using a transformation $R$ and a random restriction $\rho$ into a new $\Res(O(k))$-refutation which is practically equivalent to $\pi$ restricted by $\rho$, but which contains only DNFs with small height and thus can be simulated by a \Res-refutation of a small width. The small height/width will allow the adversary to find a partial solution of $A^-\overline{x}=b^-$ and respond to any formula from the original proof $\pi$ so that each axiom of $\tau_{W}(A,b)$ is assigned 1 and the consistency w.r.t. $\Res(k)$-derivation rules is maintained (implying that $\pi$ will not catch the adversary). We proceed with a more formal description.
\medskip

\noindent {\bf Moving to protected DNFs.} Given a $k$-DNF $F$ from $\pi$, the adversary first sets all variables in $F$ which do not appear in $\tau_{W}(A,b)$ to 0 and applies the transformation $R$ from the proof of \cite[Claim 5.8]{Rkdnf} which produces a `protected' $O(k(d+2))$-DNF $R(F)$. However, our adversary adjusts the parameters in order to accommodate a more `sparse' $\rho$. Specifically, in \cite[Definition 5.3]{Rkdnf} in Item 2, $3d$ is increased to $5d$ and in Item 3, $20kd$ is increased to $32kd$.\footnote{For clarification, we remark that $L_{i,1}, R_{i,1}$ in Item 1 of \cite[Definition 5.3]{Rkdnf} denote $L_{i1},R_{i1}$ and $Conv(L_{i\nu},R_{i\nu})$ in Item 3 is a subset of $J_i(A)$.} Further, in \cite[Definition 5.4]{Rkdnf}, $6d$ is increased to $10d$.\footnote{Again, in order to avoid a potential ambiguity, we stress that intervals $(r_{i,\nu},r_{i,\nu+1}]$ in \cite[Definition 5.4]{Rkdnf} are subsets of $J_i(A)$.} Consequently, $14kd$ in \cite[Claim 5.5 a)]{Rkdnf} becomes $22kd$. In the proof of \cite[Claim 5.7]{Rkdnf}, $3d$ becomes $5d$ and $6kd$ becomes $10kd$. Recall also that $d$ in \cite[Definitions 5.3 \& 5.4]{Rkdnf} is $d+2$ in our setting of parameters. 

$R$ (together with assigning 0's outside $\tau$) transforms $\pi$ into a set of $O(k(d+2))$-DNFs which can be completed into a $|\pi|2^{O(k(d+2))}$-size $\Res(O(k(d+2)))$-refutation of some $\Phi'$ such that each axiom of $\tau_{W}(A,b)$ gets converted to a DNF which logically follows 
from $\tau_{W}(A,b)$. 
The completion exists by the proof of \cite[Claim 5.8]{Rkdnf} and by the proof of \cite[Claim 5.7]{Rkdnf} (which guarantees that for each axiom $F'$ in $\tau_{W}(A,b)$, we have $F'\Leftrightarrow R(F')$ assuming $\tau_{W}(A,b)$), but our adversary will construct only a part of it --- we postpone the details. The definition of $R$ exploits the property $(\ref{e:assumption})$, which as we already explained can be assumed to hold for the relevant submatrix $A^-$ of $A$.

The transformation $R$ can be implemented by $poly(t,2^{k(d+2)},|W|,m')$-size circuits. This follows straightforwardly by inspection of its definition. We provide $A^-$ as a non-uniform advice of the adversary. 
The occurrence of $|W|$ is enforced by the necessity of generating literals $(y_{J_i(A)}^i)^{b_i}$ (referred to as `cosmetic') as the only way for our adversary to access bits $b_i$ is to have the bits $b_{u,1}$ with $u\in W$ stored in a non-uniform advice.\footnote{For this reason we would get at best only $poly(t^{t^{\delta}},2^n,m')$-size $\Res(k)$-adversaries for $\neg\ttable(f_n,t)=\neg\lb_{W}(f_n,t)$, if $W=\{0,1\}^n$.}
\medskip

\noindent {\bf Applying a random restriction.} Next, the adversary applies the random restriction $\rho$ from \cite[Definition 5.9]{Rkdnf}, but the parameter $2d$ in \cite[Definition 5.9]{Rkdnf} is increased to $4d$ --- again, note that in our setting $d$ in \cite[Definition 5.9]{Rkdnf} becomes $d+2$. 
In more detail, if $R(F)$ contains a variable $y^i_{\Sigma}$, the adversary constructs $\tilde{Y}_i$ and a $poly(t_0)$-size part of $J$ needed to determine $Y_i$ (the part of $J$ in $J_i(A)$). After determining $Y_i$, the adversary constructs the part of $\rho$ needed to determine $R(F)|_{\rho}$ (using bits $b_{u,1}$ again to assign the cosmetic variable $y^i_{J_i(A)}$). Similarly, if $R(F)$ contains a variable $x_j$. The fact that circuits $C_{i,j}$ from our adversary share a random string is used to ensure that in future rounds our adversary will 
re-produce the same $\rho$ which was used in previous rounds. 

In the definition of sparsity, \cite[Definition 5.1]{Rkdnf}, $2d$ is increased to $4d$. Then, by \cite[Claim 5.10]{Rkdnf}, the probability that $\rho$ is sparse is $\ge 1/2$, but the same proof shows that the probability is $\ge 1-(1-\gamma)/2$, if $C'_{\gamma}$ is big enough. Here we use $(\ref{e:assumption})$ again. To see that the proof of \cite[Claim 5.10]{Rkdnf} goes through even with our adjusted parameters, it suffices to implement the following changes in it: $9d$ becomes $18d$, $s\ge C\log m/9$ becomes $s\ge C'_{\gamma}\log m/18$, $4d$ becomes $8d$ and $|\Delta^c_v|=d$ becomes $|\Delta^c_v|=2d$.

As explained in the proof of \cite[Claim 5.2]{Rkdnf}, for any sparse $\rho$, $\tau_W(A,b)|_{\rho}$ can be transformed using a variable substitution $\rho'$ into clauses encoding a linear system of equations $A^+\overline{x}=b^+$, where $A^+$ is an $(r,d+2)$-lossless expander with $|J_i(A^+)|\ge 4(d+2)$. (If we did not adjust the parameters above, we would get only $|J_i(A^+)|\ge 2(d+2)$, which would not suffice to prove Lemma \ref{l:k1} guarantying the existence of suitable partial solutions of $A^+\overline{x}=b^+$.) 

\cite[Claim 5.14]{Rkdnf} remains valid for our adjusted parameters as well. To see that, it suffices to increase $6d$ to $10d$ in the proof of \cite[Claim 5.12]{Rkdnf} and $2d$ to $4d$ in the proof of \cite[Claim 5.13]{Rkdnf}. Therefore, by \cite[Claim 5.14]{Rkdnf}, $\rho$ collapses the height of $R(F)$ with high probability, $$\Pr_{\rho}[h(R(F)|_{\rho})\ge h]\le 1/e^{h/2^{O(k(d+2))}},$$ where $h\colone r/(C'_{h}k(d+2))$, for a big enough constant $C'_h$. Since $4|\pi|\le (1-\gamma)2^{h/2^{O(k(d+2))}}/2$, 
 with probability $\ge 1-(1-\gamma)/2$, every $F\in \pi\cup\pi'$ was transformed to a formula $R(F)|_{\rho}$ with height $\le h$. Here, $\pi'$ is an auxiliary set of $k$-DNFs of size $|\pi'|\le 3|\pi|$ which will be specified later. The 
transformation $R$ will be defined on $\pi'$.

The last three paragraphs imply that with probability $\ge \gamma$, the applications of $R$ (with assigning 0s outside $\tau$) and $\rho$ transform $\pi$ into (a set of formulas that can be completed to) a $\Res(O(k(d+2)))$-refutation of some $\Phi''$ such that each axiom of $\tau_{W}(A,b)|_{\rho}$ is converted to a DNF which follows 
from axioms which can be further interpreted (using $\rho'$) as axioms encoding $A^+\overline{x}=b^+$, where $A^+$ is an $(r,d+2)$-lossless expander satisfying $|J_i(A^+)|\ge 4(d+2)$, for all $i\in [|b^+|]$. Additionally, $\forall F\in \pi\cup \pi'$, $h(R(F)|_{\rho})\le h$. 
\bigskip

In the rest of the proof we define the responses of the adversary in the case that $R$ and $\rho$ ensured the properties above. If this does not happen and the adversary is led to respond to an axiom of $\tau_W(A,b)$ by 0, it recognizes that the transformation failed and (despite failing in this case) overwrites its response to 1 in order to ensure that it is {\em for $\tau_W(A,b)$}.
\bigskip

\noindent {\bf Finding partial solutions.} 
Our adversary already produced $R(F)|_{\rho}$, for a given $k$-DNF $F$. 
To complete the proof of Theorem \ref{t:efadv-r} it remains to explain how the adversary solves the system of equations corresponding to $R(F)|_{\rho}$ in a consistent fashion (so that it responds to each axiom of $\tau_{W}(A,b)$ by 1 and preserves the $\Res(k)$-derivation rules). For this purpose we adapt a proof of a width lower bound presented by Kraj\'i\v{c}ek \cite[Lemma 13.4.5]{Kpc}. Our proof differs from Kraj\'{i}\v{c}ek's presentation in that 
instead of using $(r,\cdot )$-expansion we use $(r,d+2)$-lossless expansion.
\bigskip

For a set of rows $I\subseteq [|b^+|]$, let $A^+_I$ be the 
matrix obtained from $A^+$ by deleting all rows in $I$ and all columns in $\bigcup_{i\in I} J_i$, where $J_i\colone J_i(A^+)$.

\begin{lemma}\label{l:k1} Let $A^+$ be an $(r,d+2)$-lossless expander as above satisfying $|J_i|\ge 4(d+2)$, for all $i\in [|b^+|]$. For any set of rows $I\subseteq [|b^+|]$ of size $|I|\le r/2$, there is $\hat{I}\supseteq I$, $|\hat{I}|\le 2|I|$ such that \begin{equation}\label{e:exp}\text{For each }i\notin \hat{I},\ \left| J_i\backslash\bigcup_{u\in \hat{I}}J_u\right|\ge |J_i|-2(d+2).\end{equation} Moreover, for any $\hat{I}$ 
satisfying $(\ref{e:exp})$, for any nonempty set $K$ of $\le r$ rows in $A^+_{\hat{I}}$, $|\partial _{A^+_{\hat{I}}}(K)|>0$.
\end{lemma}

\proof Assume $|I|\le r/2$. Put $I_0\colone I$. Add in consecutive steps $t=0,\dots$ to $I_t$ any one row $i$ as long as 
\begin{equation}\label{e:exp1}\left|J_i\cap \bigcup_{u\in I_t} J_u\right|>2(d+2).\end{equation} We claim that the process stops before $t$ reaches $|I|$. Assume not, and let $I'\colone I_{|I|}$. Then by (\ref{e:exp1}), $$\sum_{i\in I'} |J_i|-|\partial_{A^+}(I')|>\sum_{i\in I'} |J_i|-\left(\sum_{i\in I} |J_i|+\sum_{i\in I'\backslash I} (|J_i|-2(d+2))\right)=\sum_{i\in I'\backslash I} 2(d+2)=(d+2)|I'|.$$ This contradicts the expansion property of $A^+$.

Let $\hat{I}$ be the last $I_t$ in the process, so $t<r/2$, $|\hat{I}|\le 2|I|$ and $\hat{I}$ has the property (\ref{e:exp}).

For the `moreover' part, note that $$\partial_{A^+_{\hat{I}}}(K)=\partial_{A^+}(K)\backslash\bigcup_{i\in\hat{I}}J_i.$$ As for all $i\in K\backslash \hat{I}$ we have $|J_i\cap \bigcup_{u\in \hat{I}}J_u|\le 2(d+2)$, this implies that $$|\partial_{A^+_{\hat{I}}}(K)|\ge |\partial_{A^+}(K)|-\sum_{i\in K} 2(d+2)\ge\sum_{i\in K} |J_i|-(d+2)|K|-2(d+2)|K|>0.$$ The last inequality uses $|J_i|\ge 4(d+2)$.
  \qed

\medskip

Any $I\subseteq [|b^+|]$ satisfying (\ref{e:exp}) from Lemma \ref{l:k1} will be referred to as a {\em safe set of rows}. A partial assignment $\alpha$ to $\overline{x}$ assigning to 0 or 1 variables from a set $dom(\alpha)$ is {\em safe with support} $I$ if $\{j\mid x_j\in dom(\alpha)\}\subseteq\bigcup_{i\in I} J_i$ for some safe $I$. 
A safe partial assignment $\alpha$ with support $I$ 
is a {\em safe partial solution} of $A^+x=b^+$ {\em with support $I$} if and only if $\bigoplus_{j\in J_i}\alpha(x_j)=b_i^+$, for all $i\in I$.

\begin{lemma}\label{l:k2} Assume $I\subseteq I'\subseteq [|b^+|]$ are safe sets and $|I'\backslash I|\le r$. Assume further that $\alpha$ is a safe assignment with support $I$, 
and that $c_i\in\{0,1\}$, for $i\in I'\backslash I$. 

Then the assignment $\alpha$ can be extended to a safe assignment $\alpha'$ with support $I'$
such that $$\bigoplus_{j\in J_i} \alpha'(x_j)=c_i,$$ for all $i\in I'\backslash I$.
\end{lemma}

\proof By Lemma \ref{l:k1}, each subset of $I'\backslash I$ has a non-empty border in $A^+_I$. In particular, no subset of $I'\backslash I$ can be a linearly dependent set of rows of $A^+_I$. Hence the system $$\bigoplus_{j\in J_i\backslash dom(\alpha)}x_j=c_i\oplus \bigoplus_{j\in J_i\cap dom(\alpha)} \alpha(x_j),\text{ for }i\in I'\backslash I,$$ has a solution $\theta$. Put $\alpha'\colone\alpha\cup\theta$.\qed 
\medskip

The last lemma we need is a result of Segerlind, Buss and Impagliazzo \cite[Theorem 5.1]{SBI} which translates $\Res(k)$-derivations with lines of small height to \Res-derivations of small width. Our formulation is slightly more general but it follows directly from the proof presented in \cite{SBI}. Given a $k'$-DNF $M$ of height $h(M)\le h'$, denote by ${\cal M}^{h'}_M$ the set of clauses $M'$ of width $h'$ built on variables of $M$ such that $\neg M'$ falsifies all terms in $M$.

\begin{lemma}[\cite{SBI}]\label{l:sbi} If there is a $\Res(k')$-derivation of a $k'$-DNF $M_0$ that has height $h(M_0)\le h'$ from a set of $k'$-DNFs $\cal M$ such that each $M_1\in {\cal M}$ has height $h(M_1)\le h'$ and each $k'$-DNF $M$ in the derivation has height $h(M)\le h''$, where $h''\ge h'$, then every clause $M_0'\in {\cal M}^{h'}_{M_0}$ can be derived from $\bigcup_{M_1\in {\cal M}}{\cal M}^{h'}_{M_1}$ by a \Res-derivation of width $\le h''k'$ (possibly using the weakening rule).
\end{lemma}
\bigskip

We are finally ready to finish the proof of Theorem \ref{t:efadv-r}.

In each round $j\ge 1$, given the $j$th $k$-DNF $F$, our adversary will construct an $h$-CNF $\overline{F}\colone {\cal M}_{R(F)|_{\rho}}^h$ equivalent to $R(F)|_{\rho}$ by going through all possible clauses $M$ of width $h$ and adding $M$ to $\overline{F}$ if $\neg M$ falsifies all terms in $R(F)|_{\rho}$. The $h$-CNF exists and the construction succeeds because $h(R(F)|_{\rho})\le h$. Since there are only $poly(t,|W|)$ variables in $R(F)|_{\rho}$ (recall that all variables outside $\tau_W(A,b)$ are automatically set to 0), $\overline{F}$ can be constructed by a $poly((t+|W|)^h,|R(F)|_{\rho}|)$-size circuit. Further, the adversary will produce $\alpha_1,I_1,\dots,\alpha_j,I_j$, where $\alpha_{j'}$, $ j'\in [j],$ is a safe partial solution of $A^+\overline{x}=b^+$ with support $I_{j'}$ of size $|I_{j'}|\le 2h$. The partial solution $\alpha_j$ will be used to evaluate $F$: either a clause in $\overline{F}$ will be falsified by $\overline{\rho\circ\alpha_j}$, in which case the response of the adversary to $F$ is 0, or otherwise the response to $F$ is 1. Here (and in the rest of the proof) we abuse the notation and use $\rho'$ to identify variables of $R(F)$ with variables encoding $A^+\overline{x}=b^+$ --- note that the relevant part of $\rho'$ can be computed by $poly(t,|R(F)|,|W|,m')$-size circuits from $R(F)$ (the access to $m'$ ensures that the adversary can use new random bits when constructing $\rho$). The partial assignment $\overline{\rho\circ\alpha_j}$ is the extension of $\rho\circ\alpha_j$ which assigns also the variables $y^i_{\Sigma}$ in rows $I_j$ whose values are determined by $\rho\circ\alpha_j$ and $A^+\overline{x}=b^+$.
\smallskip

\noindent {\em Response to axioms.} Let us first observe that the response of such an adversary to each axiom $F$ in $\tau_{W}(A,b)$ in each round $j\ge 1$ will be 1. Denote the system of equations of $A^+\overline{x}=b^+$ corresponding to rows from a set $I\subseteq [|b^+|]$ by $A^+\overline{x}=b^+|_I$. Let $I^+$ be a minimal set of rows in $[|b^+|]$ such that all variables of $R(F)|_{\rho}$ are in the clauses encoding the equations $A^+\overline{x}=b^+|_{I^+}$. As $F$ has a constant width, \cite[Claim 5.7 a)]{Rkdnf} implies that $|I^+|\le O(k(d+2))$ 
and $|I_j\cup I^{+}|\le 2h+O(k(d+2))\le r/2$. By Lemma \ref{l:k1}, there is thus an extension of $I_j\cup I^+$ to a safe set of size $\le r$ and by Lemma \ref{l:k2} there is a solution $\alpha\supseteq \alpha_j$ of $A^+\overline{x}=b^{+}|_{I_j\cup I^+}$. A closer look at the definition reveals that $R(F)$ results from rewriting some variables $y^i_{\Sigma^i_\nu}$ in $F$ by linear forms in $x_{\ell}$'s which these variables $y^i_{\Sigma^i_\nu}$ represent according to the $i$'th equation in $\tau_{W}(A,b)$ and the sign of the `anchor' variable. It is not hard to check that $A^+\overline{x}=b^{+}|_{I^+}$ implies that the $y^i_{\Sigma^i_\nu}$'s (whose assignment is now determined by $\rho\circ\alpha$) are equivalent to these linear forms. More precisely, if $y^i_{\Sigma^i_{\nu_{i\beta}}}$ is the anchor variable and $\nu>\nu_{i\beta}$, then $A^+\overline{x}=b^+|_{I^+}$ implies \begin{align*}y^i_{\Sigma^i_\nu}\oplus\bigoplus_{j\in \Sigma^i_{\nu_+}\backslash \Sigma^i_\nu} x_j=\rho(y^i_{\Sigma^i_{\nu_+}}) & =\rho(y^i_{\Sigma^i_{\nu_-}})\oplus \bigoplus_{j\in (\Sigma^i_{\nu_+}\backslash\Sigma^i_{\nu_-}) \cap J_x(\rho)} \rho(x_j)\oplus\bigoplus_{v\in (\nu-,\nu+]} (b|_{\rho})_{i,v}\\
& =y^i_{\Sigma^i_{\nu_{i\beta}}}\oplus\bigoplus_{j\in \Sigma^i_{\nu_+}\backslash \Sigma^i_{\nu_{i\beta}}} x_j,\end{align*} where $\Sigma^i_\nu\subseteq\Sigma^i_{\nu_+}$ is the minimal initial segment of $J_i(A)$ such that $y^i_{\Sigma^i_{\nu_+}}\in sup(\rho)$ and $\Sigma^i_{\nu_-}\subseteq\Sigma^i_{\nu_{i\beta}}$ is the maximal initial segment of $J_i(A)$ such that $y^i_{\Sigma^i_{\nu_-}}\in sup(\rho)$, so $y^i_{\Sigma^i_\nu}=y^i_{\Sigma^i_{\nu_{i\beta}}}\oplus\bigoplus_{j\in \Sigma^i_\nu\backslash \Sigma^i_{\nu_{i\beta}}} x_j$. If $\nu<\nu_{i\beta}$, we get the same equations but $\Sigma^i_{\nu_+}$ is defined so that $\Sigma^i_{\nu_{i\beta}}\subseteq\Sigma^i_{\nu_+}$ and $\Sigma^i_{\nu_-}$ so that $\Sigma^i_{\nu_{-}}\subseteq\Sigma^i_{\nu}$, which yields $y^i_{\Sigma^i_\nu}=y^i_{\Sigma^i_{\nu_{i\beta}}}\oplus \bigoplus_{j\in \Sigma^i_{\nu_{i\beta}}\backslash \Sigma^i_{\nu}}x_j$. Since $\overline{\rho\circ\alpha}$ satisfies $F$, $R(F)|_{\overline{\rho\circ\alpha_j}}$ is satisfiable and the adversary responds to $F$ by 1.
\smallskip

\noindent {\em Round 1 and Weakening.} The first $k$-DNF our adversary receives is the empty clause and $\alpha_1\colone\emptyset, I_1\colone\emptyset$. Recall the derivation rules of $\Res(k)$ from \cite[Definition 2.6]{Rkdnf}. If the $j$th DNF $F$ is a premise of an application of the weakening rule whose conclusion is a DNF $F_{j'}$ to which the adversary responded by 0 in round $j'<j$, the adversary recognizes that $F_{j'}$ is derived using the weakening rule or the CUT-rule. Even if the rule is CUT, it pretends that it is Weakening and sets $\alpha_j\colone\alpha_{j'}, I_j\colone I_{j'}$. As every clause in $\overline{F_{j'}}$ restricted to the variables of $\overline{F}$ is in $\overline{F}$, this 
guarantees the consistency w.r.t. the rule. 
\smallskip

\noindent {\em Round $j>1$ for CUT, AND-elimination and AND-introduction.} Suppose that the $j$th $k$-DNF $F\vee T$ is a premise of the CUT-rule, AND-elimination or AND-introduction whose conclusion is a DNF $F\vee G$ to which the adversary responded by 0 in round $j'<j$. Further, assume that the situation is not consistent with an application of the weakening rule. If the situation is consistent with an application of AND-elimination, the adversary pretends that the rule is AND-elimination. 
 In the remaining cases, the adversary does not necessarily recognize if $F\vee G$ is derived using CUT or AND-introduction. If this happens, it simulates both alternatives. 
The adversary does not necessarily recognize all premises of the derivation step either. In the case of AND-introduction this can happen if $F\vee G=F$, so they simulate all possible applications of AND-introduction that are consistent with the given formulas. Since the size of terms in $F$ is $\le k$, this yields $\le t^{O(ak)}$ simulations. When simulating CUT with a principal formula $T'$, the adversary pretends that the premises are $F\vee G\vee T'$ and $F\vee G\vee\neg T'$. Since the adversary might not recognize the principal formula that is being cut in this step, they simulate all possible applications of the CUT rule that are consistent with the situation. This yields $\le t^{O(ak)}$ simulations. The transformation $R$ is defined on all formulas $F\vee G\vee T'$, $F\vee G\vee \neg T'$ in these simulations. All formulas $F\vee G\vee T'$ and $F\vee G\vee \neg T'$ in a simulation of a CUT in $\pi$ such that $T'$ is actually the principal formula of the CUT in $\pi$ are added to $\pi'$. This adds to $\pi'$ at most two new formulas for each CUT in $\pi$, i.e. $\le 2|\pi|$ formulas in total. The adversary does not need to construct $\pi'$. Denote the set of the premises of the simulated rule (as pretended by the adversary) by $\Gamma$. 
\def\part{As $|J_u(A^+)\cap J_{u'}(A^+)|>1$ only if $u,u'$ are from the same copy of $A_0$ (in the construction of $A^+$), the algorithm from the proof of Lemma \ref{l:k1} producing $\hat{I}_{j',i}$ can be implemented by a circuit of size $poly(t_0)$. By Lemma \ref{l:k2} there exists a solution $\hat{\alpha}_{j',i}\supseteq\alpha_{j',i}$ of $A^+\overline{x}=b^+|_{\hat{I}_{j',i}}$. Our adversary finds $\hat{\alpha}_{j',i}$ efficiently using Gaussian elimination (and the bits $b_{u,1}$ with $u\in W$ stored in the non-uniform advice). The adversary then proceeds to the predecessor $M_{i+1}\in\Pi'$ of $M_i$ falsified by $\overline{\rho\circ\hat{\alpha}_{j',i}}$.}

As explained in the proof of \cite[Claim 5.8]{Rkdnf}, the $R$-image of the $\Res(k)$-derivation step concluding $F\vee G$ 
can be completed to a $\Res(O(k(d+2)))$-derivation $\Pi$ using the axioms of $\tau_{W}(A,b)$.
\footnote{Note that terms $R((y^i_{J_i(A)})^{b_i})$, which are by definition empty (by the proof of \cite[Claim 5.8]{Rkdnf} $R$ is defined on literals and by the proof of \cite[Claim 5.7 a)]{Rkdnf} cosmetic literals are removed even if they appeared in the original term), are treated as valid and efficiently derivable. We prefer to set $R(\bigwedge_{i\in I}(y^i_{J_i(A)})^{b_i}\wedge\bigwedge_{i'\in I'}\neg y^{i'}_{\emptyset})\colone \bigwedge_{i\in I}(y^i_{J_i(A)})^{b_i}\wedge\bigwedge_{i'\in I'} \neg y^{i'}_{\emptyset}$, for $I,I'\subseteq [m]$ such that $I\cup I'\ne \emptyset$. Formally, this is a new transformation $R'$ which might not output a protected DNF, but whenever $R(F)\ne R'(F)$, $R'(F)|_{\rho}$ is trivialized because $\rho((y^i_{J_i(A)})^{b_i})=\rho(\neg y^i_{\emptyset})=1$.} 
Now we finish the specification of $\pi'$ by including in it also the non-principal formulas of each derivation step in $\pi$ except CUTs. 
This adds to $\pi'$ at most $|\pi|$ formulas, making the total size of $\pi'$ at most $|\pi'|\le 3|\pi|$. Given that the height of the $R|_{\rho}$-image of the non-principal formula of the simulated rule is $\le h$ (which is the case if the simulated derivation step is in $\pi$ or if the simulated rule is CUT), it is easy to check that height of each formula in $\Pi|_{\rho}$ is $\le h+O(k(d+2))$. Hence, by Lemma \ref{l:sbi}, the clause in $\overline{F\vee G}$ falsified by $\overline{\rho\circ\alpha_{j'}}$ can be derived from $\overline{\Phi_{\Gamma}}$'s, for $\Phi_{\Gamma}\in\Gamma$, and from $\tau_W(A,b)|_{\rho}$ by a \Res-derivation $\Pi'$ of width $w\le O(k(d+2)h+(k(d+2))^2)\le O(k(d+2)h)$. The adversary can construct $\Pi'$ by listing all possible clauses of width $\le w$ (there are $poly((t+|W|)^w)$ of them) and gradually deriving every clause of width $\le w$ that can be derived from $\overline{\Phi_{\Gamma}}$'s, for $\Phi_{\Gamma}\in\Gamma$, and from $\tau_W(A,b)|_{\rho}$ in $poly((t+|W|)^w)$ steps. If the adversary does not succeed in constructing $\Pi'$ because the assumption on the height is not met, the particular branch of the simulation can be aborted.

Next, the adversary simulates its evaluation on $\Pi'$. In each step $i$ of the evaluation, it stands on a clause $M_i\in\Pi'$ of width $\le w$ and holds a safe partial solution $\alpha_{j',i}$ of $A^+\overline{x}=b^+$ with a support $I_{j',i}$ of size $|I_{j',i}|\le 2w$ such that $\overline{\rho\circ\alpha_{j',i}}$ falsifies $M_i$. Initially $\alpha_{j',0}\colone\alpha_{j'}$, $I_{j',0}\colone I_{j'}$ and $M_0$ is the clause in $\overline{F\vee G}$ falsified by $\overline{\rho\circ\alpha_{j'}}$. If the adversary wants to move to a predecessor of $M_i$ in $\Pi'$, it identifies the variable $\ell$ which was resolved in the step in $\Pi'$ concluding $M_i$ and a row $r_0$ in $A^+\overline{x}=b^+$ such that $\ell$ occurs in the clauses encoding $A^+\overline{x}=b^+|_{\{r_0\}}$. Since $|I_{j',i}|+1\le r/2$, for a sufficiently big $C'_{h}$, by Lemma \ref{l:k1}, there exists a safe set of rows $\hat{I}_{j',i}\supseteq I_{j',i}\cup \{r_0\}$ of size $|\hat{I}_{j',i}|\le r$. \part


Since $M_{i+1}$ has width $\le w$, there exists a set $I^w_{j',i}\subseteq \hat{I}_{j',i}$ of $w$ rows such that $\overline{\rho\circ\beta'}$ falsifies $M_{i+1}$, where $\beta'$ is the restriction of $\hat{\alpha}_{j',i}$ to the variables with indices in $\bigcup_{i\in I^w_{j',i}}J_i$. 
The adversary can find $I^w_{j',i}$ efficiently. 
Finally, the adversary constructs the safe set of rows $I_{j',i+1}\supseteq I^w_{j',i}$, $I_{j',i+1}\subseteq \hat{I}_{j',i}$, of size $|I_{j',i+1}|\le 2w$ using the algorithm from the proof of Lemma \ref{l:k1} again and sets $\alpha_{j',i+1}$ to be the restriction of $\hat{\alpha}_{j',i}$ to the variables with indices in $\bigcup_{i\in I_{j',i+1}} J_i$. 
If $M_{i+1}$ is an initial clause of $\Pi'$, it sets $\alpha_{j}\colone\alpha_{j',i+1}$ and $I_{j}\colone I_{j',i+1}$. Note that in such case, $|I_{j}|\le 2h$ can be ensured. 

The adversary cannot arrive in an axiom of $\tau_W(A,b)$, it arrives in a clause of $\overline{\Phi_{\Gamma}}$, for some $\Phi_{\Gamma}\in\Gamma$. Hence, if the simulated rule was AND-elimination, the adversary responds by 0 to $F\vee T$ and remains consistent. If the rule was CUT (but not consistent with Weakening) or AND-introduction and in at least one of these cases the simulation ended up in a clause containing a clause of $\overline{F\vee T}$, the adversary outputs 0. The simulation of a rule is independent of the premise received by the adversary (except that some branches of the simulation might abort earlier or not even start) and in the case of CUT we have that every clause in $\overline{F\vee G\vee T}$ restricted to the variables of $\overline{F\vee T}$ is in $\overline{F\vee T}$. Therefore, the adversary responds by 0 to at least one of the premises  and remains consistent w.r.t. every rule. 
\smallskip


\noindent {\em Efficiency.} The least inefficient part of the algorithm implemented by our adversary was the construction of the \Res-derivation $\Pi'$ which took $poly(t^{rad},m')$ steps
. In each round $j$, the adversary uses a circuit of size $j\cdot poly(t^{rad},m')\le m'\cdot poly(t^{rad},m')$ to produce $\alpha_{j'}, I_{j'}$, for all $j'\in [j]$. The complexity of the adversary is thus $poly(t^{rad},m')$ and the proof of Theorem \ref{t:efadv-r} is completed.

\subsection{Pseudowidth}\label{ssec:pwd}



We present a constructive version of the lower bound for the weak pigeonhole principle $\PHP^m_n$ based on the method of pseudowidth, cf. \cite{Rwphp,Rpw1,Rpw2,Rpw3,dRpw}. 
The lower bound proceeds in two steps. In the first step, a supposed short \Res refutation of $\PHP^m_n$ is transformed into a \Res refutation of small pseudowidth. In the second step, it is shown that the resulting small pseudowidth \Res refutation cannot exist. In order to make the lower bound constructive, we modify the second step. 
\bigskip

$\neg \PHP^m_n$ is a CNF formula using variables $x_{i,j}$, $i\in [m], j\in [n]$. The intended meaning of $x_{i,j}$ is that the pigeon $i$ is mapped to the hole $j$. $\neg \PHP^m_n$ consists of the following clauses:
\begin{itemize}

\item[] (pigeon axioms) $\bigvee_{j\in [n]} x_{i,j}$, for $i\in [m]$;

\item[] (hole axioms) $\neg x_{i,j}\vee \neg x_{i',j}$, for $i\ne i'\in [m], j\in [n]$.

\end{itemize}

Given a clause $C$, define a clause $C^{{\sf mon}}$ by replacing each negated literal $\neg x_{i,j}$ in $C$ by $\bigvee_{i'\ne i} x_{i',j}$.

For a clause $C$ and pigeon $i$, $$N_{C}(i)\colone\{j\in [n] \mid C^{\sf mon}\text{ is satisfied if }x_{i,j}=1\}.$$
Further, define $\Delta_{C}(i)\colone|N_C(i)|$. 

Let $d=(d_1,\dots,d_m)$ 
be a vector such that $d_i\in \mathbb{N}\setminus \{0\}, \delta\in\mathbb{R}$ and $d_i>\delta$ for all $i\in [m]$. Then \begin{align*}P_d(C)\hspace{2pt}\colone\ & \{i\in [m]\mid \Delta_C(i)\ge d_i\},\\  P_{d,\delta}(C)\hspace{2pt}\colone\ & \{i\in [m]\mid \Delta_C(i)\ge d_i-\delta\}.\end{align*} The pseudowidth of $C$ is $w_{d,\delta}(C)\colone|P_{d,\delta}(C)|$. For a set of clauses ${\cal A}$ and a \Res refutation $\pi$, the pseudowidth of $\pi\setminus{\cal A}$ is $w_{d,\delta}(\pi\setminus {\cal A})\colone max_{C\in \pi\wedge C\not\in {\cal A}} w_{d,\delta}(C)$. A clause $C$ such that $|P_{d}(C)|\ge w_0$ is referred to as a {\em $(w_0,d)$-subaxiom}.

In this section, we include the weakening rule in \Res, i.e. a clause $C$ can be derived from a clause $D$ if $D\subseteq C$. We denote the number of clauses in a \Res refutation $\pi$ by $L_{\pi}$.
\bigskip

The first step of the lower bound based on the pseudowidth method is captured in the following lemma. We formulate it in a stronger way than usual: Instead of concluding the existence of a vector $d$, we claim that a random vector has the desired properties (even with our {\em monotonized} definition of pseudo-width). This is, in fact, what the standard proof of the lemma shows, cf. \cite[Lemma 4.2]{dRpw}.

\begin{lemma}[Pseudowidth reduction]\label{l:pwub} Let $\delta\colone n/\log m>1$, $3<m,n\in\mathbb{N}$. There is a distribution of $m$-dimensional integer vectors ${\bm d}$ depending only on $m,n$ such that each vector $d=(d_1,\dots,d_m)$ sampled according to ${\bm d}$ satisfies $\delta<d_i\le n$, for all $i\in [m]$, and such that for each \Res refutation $\pi$ of a set of clauses $\Phi$ built on variables of $\PHP^m_n$, for each $w_0\in [m]$ and a random vector $d=(d_1,\dots,d_m)$ sampled according to ${\bm d}$, with probability $\ge 1-1/L_{\pi}$, we have a set ${\cal A}$ of $(w_0,d)$-subaxioms with $|{\cal A}|\le L_{\pi}$ and a \Res refutation $\pi'\subseteq \pi$ of $\Phi\cup{\cal A}$ satisfying $$w_{d,\delta}(\pi'\setminus {\cal A})=O(w_0+\log L_{\pi}).$$ 
\end{lemma}

\begin{lemma}[Pseudowidth lower bound]\label{l:pwlb} Let $m,n,r\in \mathbb{N}\setminus\{0\}$ and $4\le\delta\in \mathbb{R}$. Suppose that $d=(d_1,\dots,d_m)$ is an integer vector such that $\delta<d_i\le n$, for all $i\in [m]$. Further, let 
$\cal A$ be an arbitrary set of $(1,d)$-subaxioms such that ${\cal A}\cap \neg\PHP^m_n=\emptyset$. 
Then for every \Res refutation $\pi$ of $\neg \PHP^m_n\cup{\cal A}$ satisfying $L_{\pi}< 2^{r\delta/8n}$, we have 
$$w_{d,\delta}(\pi\setminus {\cal A})>\delta/6r.$$
\end{lemma}

\proof For the sake of contradiction assume that $\pi$ is a \Res refutation of $\neg \PHP^m_n\cup {\cal A}$ with $w_{d,\delta}(\pi\setminus {\cal A})\le \delta/6r$. W.l.o.g. $|{\cal A}|\le L_{\pi}$ since all unused $(1,d)$-subaxioms can be erased from ${\cal A}$. For a simplification of notation we assume that for all $C\in {\cal A}$ we have $P_{d,\delta}(C)=P_{d}(C)$ and $P_d(C)$ consists of a single pigeon. (If there are more pigeons in $P_{d,\delta}(C)$, we choose one from $P_{d}(C)$ in an arbitrary way.) 

To reach a contradiction we trace $\pi$ from its last clause to an axiom. During the trace, when we are at a clause $C$, we maintain a partial multi-matching $\rho_C$ on $P_{d,\delta}(C)$ which does not satisfy $C^{\sf mon}$. For each pigeon $i$ from the domain of $\rho_C$, $\rho_C$ sets $x_{i,j}=1$ for $r$ not necessarily different $j$'s and $x_{i,j}=0$ for all the remaining $j$'s. Additionally, for all $i\ne i'$ in the domain of $\rho_C$, for all $j$, $\rho_C$ sets $x_{i,j}=0$ or $x_{i',j}=0$. Each $\rho_C$ will be generated using a randomized procedure.

The construction will ensure that the trace will never reach a hole axiom.
If we reach a pigeon axiom $C$ of the form 
$\bigvee_j x_{i,j}$, then $i\in P_{d,\delta}(C)$, so $\rho_C$ maps $i$ into some hole and $C^{\sf mon}$ is satisfied.
%
The desired contradiction will be obtained by showing that with high probability over the randomness used to generate $\rho_C$, $C^{\sf mon}$ for $C\in {\cal A}$ is satisfied by $\rho_C$.
\medskip

Denote the assignment that we have at the $\ell$th step by $\rho_{\ell}$. Initially, $\rho_1$ is empty. Assume that in step $\ell\ge 1$ we have $\rho_{\ell}=\rho_C$ and that $C_0,C_1$ are the predecessors of $C$ in $\pi$, i.e. $C$ is derived by resolving on $C_0,C_1$. First, we extend $\rho_{\ell}$ to $\rho'_{\ell}$ in the following way:
\begin{itemize}
\item[$\circ$] For each $i\in (P_{d,\delta}(C_0)\cup P_{d,\delta}(C_1))\setminus P_{d,\delta}(C)$, choose $r$ (not necessarily different) holes $j_1,\dots,j_r$ uniformly at random from $[n]\setminus (N_{C}(i)\cup R)$, map $i$ to $j_1,\dots,j_r$ (i.e. set $x_{i,j_1},\dots,x_{i,j_r}$ to 1 and the remaining $x_{i,j}$'s to 0) and add $j_1,\dots,j_r$ to $R$. Here, $R$ is initially the set of holes occupied by pigeons assigned by $\rho_{\ell}$.
\end{itemize}

\noindent During the construction of $\rho'_{\ell}$, the size of $R$ does not exceed $3wr\le \delta/2$, where $w\colone w_{d,\delta}(\pi\setminus {\cal A})\ge 1$.\footnote{We have $w\ge 1$ because either a pigeon axiom is in $\pi$ or 
there is a clause in $\pi$ derived from a clause in ${\cal A}$ by weakening or by cut. In all cases it follows that $w\ge 1$.} Moreover, as $i\notin P_{d,\delta}(C)$, we have $\Delta_C(i)<d_i-\delta$, which leaves at least $n-3wr-d_i+\delta\ge 1$ free holes for each pigeon. So the construction of $\rho'_{\ell}$ is well-defined. Since $i$ was mapped to $j$'s such that $j\not\in N_C(i)$, the resulting $\rho'_{\ell}$ does not satisfy $C^{\sf mon}$ and thus it does not satisfy at least one of the clauses $C_0^{\sf mon},C_1^{\sf mon}$. If neither of $C_0, C_1$ is a weakening of a hole axiom, we proceed to the clause which is not satisfied, w.l.o.g. it is $C_0^{\sf mon}$, and define $\rho_{\ell+1}=\rho_{C_0}$ as $\rho'_{\ell}$ restricted to $P_{d,\delta}(C_0)$. If $C_0$ or $C_1$ is a weakening of a hole axiom, say $C_1$, it can be observed that $C^{\sf mon}_0\subseteq C^{\sf mon}$ (assuming that $C$ is not a weakening of a hole axiom), so the trace can proceed to $C_0$ and we set $\rho_{\ell+1}=\rho_{C_0}$ as $\rho'_{\ell}$ restricted to $P_{d,\delta}(C_0)$. If $C$ is derived via the weakening rule, the construction of $\rho_{\ell+1}$ is analogous. 
The partial multi-matchings $\rho_{C}$ satisfy the required properties except that we need to show that with high probability, $C^{\sf mon}$ for $C\in {\cal A}$ is satisfied by $\rho_{\ell}=\rho_C$ for $\ell\le L_{\pi}$. 
\medskip

Suppose $D\in {\cal A}$. We want to bound the probability that $\exists t<\ell, E_t^D$ and $D^{\sf mon}$ is not satisfied by $\rho_{\ell}=\rho_C$ at a step ${\ell}\le L_{\pi}$, where $E_t^D$ is the event that the pigeon $i\in P_{d}(D)$ is assigned in the process of generating $\rho_{t+1}$ from $\rho_t$.
\medskip

Consider $\rho_t$ such that $E^D_{t}$ takes place. During the construction of $\rho_{{t}+1}$, 
while assigning the pigeon $i$ the $r'$th hole $j_{r'}$, $r'\le r$, we have at least $n-3wr-d_i+\delta$ possible choices among the free holes. 
Denote the set of free holes by $H^{r'}_{\mathrm{free}}$. On the other hand, since $i\in P_{d}(D)$, there are at least $d_i$ different holes such that sending $i$ to one of these holes will satisfy $D^{\sf mon}$. Denote the set by $H^{r'}_{\mathrm{sat}}$. Therefore, the probability of the event $B_{\rho_t}^{D}$ that assigning the pigeon $i$ in the process of generating $\rho_{t+1}$ from $\rho_{t}$ does not satisfy any literal in $D^{\sf mon}$ is \begin{align*}\Pr[B^{D}_{\rho_t}]= &\prod_{r'\le r}\left(1- \frac{|H^{r'}_{\mathrm{free}}\cap H^{r'}_{\mathrm{sat}}|}{|H^{r'}_{\mathrm{free}}|}\right) \le \prod_{r'\le r}\left(1-\frac{|H^{r'}_{\mathrm{free}}|+|H^{r'}_{\mathrm{sat}}|-n}{n}\right)\\ \le &\left(1- \frac{\delta-3wr}{n}\right)^r\le \left(1-\frac{\delta}{4n}\right)^r.\end{align*}

The estimation of $\Pr[B^{D}_{\rho_t}]$ is independent of $\rho_{t}$ and the assignment of other pigeons than $i$ in the process of constructing $\rho_{t+1}$ from $\rho_{t}$. Hence, for each $\ell\le L_{\pi}$, 
\begin{align*}\Pr\left[\exists t<\ell, E_t^D\cap B^{D}_t\right] \le & \sum_{t\le \ell} \left(1-\frac{\delta}{4n}\right)^r\le L_{\pi}\left(1-\frac{\delta}{4n}\right)^{r}\le L_{\pi}2^{-r\delta/4n},\end{align*}
where $B^D_t$ is defined like $B^D_{\rho_t}$ but in the probability space of all $\rho_{\ell}$'s, and 
\begin{align*}\Pr[C\in {\cal A}\wedge C^{\sf mon}\text{ is not satisfied by }\rho_C]\le& \sum_{D\in {\cal A}}\Pr[\exists t<\ell, E_t^D\cap B^{D}_t] \le |{\cal A}|L_{\pi}/2^{r\delta/4n}<1.\end{align*}
\qed 
\bigskip

Lemmas \ref{l:pwub}-\ref{l:pwlb}, with $m=\lfloor 2^{n^{1/5}}\rfloor$, $r=\lfloor n^{1/2}\rfloor $ and $\Phi=\neg \PHP^m_n$, imply that there is no \Res refutation of $\PHP^m_n$ of size $2^{\epsilon n^{1/4}}$, for a sufficiently small $\epsilon$.
\bigskip

An advantage of our \Res lower bound in comparison to \cite{Rwphp,Rpw1,Rpw2,Rpw3,dRpw} 
 is that it constructs efficient \Res-adversaries for $\neg \PHP^m_n$. 

We remark that $\PHP^{|W|}_{s}$ is closely related to the 
 statement that a Boolean function $f_n$ with $n$ inputs cannot be computed by a perfect DNF (PDNF) of size $s$ on inputs from $W\subseteq\{0,1\}^n$. Here, a PDNF of size $s$ is a disjunction $\bigvee_{j\le s} K_j$, where each $K_j$ is a conjunction of $n$ literals such that each variable from $\{x_1,\dots,x_n\}$ appears in $K_j$ in the positive or negative form. 

To see the connection, interpret the variable $x_{a,j}$ of $\PHP^{|W|}_s$ as the statement $K_j(a)=1$ for $a\in W\colone f^{-1}(1)$. If we wanted to construct efficient adversaries for the encoding of PDNF lower bounds from \cite{Rpw3}, we would need to make our lower bound work for the `onto' version of the `functional' $\PHP^m_n$.

\def\naturalencoding{
$y_{a,j,k}$ encoding that the first $k\in [n]\cup \{0\}$ literals in $K_j$ are consistent with $a\in W$;

$s_{j,k}$ encoding the sign with which $x_k$ occurs in $K_j$, where $j\in [s], k\in [n]\cup\{0\}$;

$y_{a,j}$ encoding $\bigvee_{j'} K_{j'\le j}(a)$, where $j\in [s], a\in W$;

$q_j$ encoding whether $K_j$ is empty or not, where $j\in [s]$.
\medskip

$\neg\lb^*_W(f_n,s)$ is obtained by expanding the following equations into a 3-CNF:
\begin{itemize}
\item[] $y_{a,j,k}\equiv y_{a,j,k-1}\wedge s^{a_k}_{j,k}$ (with $y_{a,j,0}\coloneq_j$);

\item[] $y_{a,j}\equiv y_{a,j-1}\vee y_{a,j,n}$ (with $y_{a,0}\colone0$);

\item[] $y_{a,s}\equiv f_n(a)$, for $a\in W$.
\end{itemize}
}


\begin{theorem}[Efficient adversary for the pseudowidth lower bound]\label{t:epw}\ \newline
Let $\gamma<1$ be an arbitrary constant. There is $\epsilon>0$ and $n_0\in \mathbb{N}$ such that for all $n\ge n_0$ there is a $poly(m,m')$-size randomized \Res-adversary $C$ for $\neg \PHP^m_n$, $m=\lfloor 2^{n^{1/5}}\rfloor$, such that for each $2^{\epsilon n^{1/4}}$-size \Res refutation $\pi$ of any set of clauses, with probability $\ge \gamma$ over the randomness of $C$, $\pi$ does not catch $C$. Here, $m'$ is the input length of circuits in $C$.
\end{theorem}

The proof of Theorem \ref{t:epw} follows by combining Lemma \ref{l:pwub} with the proof of Lemma \ref{l:pwlb}. 

A complication is that when the adversary possesses $\rho_{C}$ and receives a predecessor $C_0$ of $C$, it might not know what $C_1$ is. In such case the adversary pretends that $C_1$ is $C\vee \neg (C_0\setminus C)$. More precisely, the adversary will automatically replace each clause $C_i\in\pi$, which was used in $\pi$ to derive $C$ from $C_0$ and $C_1$, by $C\vee \neg (C_{1-i}\setminus C)$. The transformation maps $\pi$ into a sequence of clauses which can be completed into a valid \Res refutation $\rho^1(\pi)$ of axioms of $\pi$: For each clause $C\vee\neg(C_{1-i}\setminus C)$ replacing a clause $C_i$, $\rho^1(\pi)$ contains a weakening step deriving $C\vee \neg(C_{1-i}\setminus C)$ from $C_i$. Since $\rho^1(\pi)$ introduces at most two new clauses for each derivation step in $\pi$, we have $L_{\rho^1(\pi)}\le 3L_{\pi}$.

Given a clause $C\in\rho^1(\pi)$, the adversary sets all variables that do not appear in $\neg \PHP^m_n$ to 0. This substitution, denoted $\rho^2$, transforms $\rho^1(\pi)$ into $\rho^2(\rho^1(\pi))$. Next, the adversary chooses a random vector $d$ according to the distribution ${\bm d}$ from Lemma \ref{l:pwub}. Since the circuits defining the adversary share a random string, the adversary can reproduce the same $d$ in each round. By Lemma \ref{l:pwub}, with probability $\ge 1-1/L_{\rho^1(\pi)}$, $\rho^2(\rho^1(\pi))$ contains a \Res-refutation $\pi'$ of $\Phi\cup {\cal A}$ such that $w_{d,\delta}(\pi'\setminus {\cal A})\le O(n^{1/4})$, where ${\cal A}$ is a set of $(1,d)$-subaxioms and $\Phi$ is a subset of axioms of $\rho^2(\rho^1(\pi))$. W.l.o.g. $1-1/L_{\rho^1(\pi)}\ge \gamma+(1-\gamma)/2$ as otherwise we can extend $\pi$ using the weakening rule on the last clause sufficiently many times. 

The adversary then simulates the construction of partial multi-matchings $\rho_{C}$ from the proof of Lemma \ref{l:pwlb} on $\pi'$, with $r=\lfloor n^{1/2}\rfloor$, and responds to $C\in\pi$ by 0 if and only if it ends up in $C$ when traversing $\pi'$. Since there might be axioms in $\pi'$ which are not in $\neg \PHP^m_n\cup{\cal A}$, we might not have $w\ge 1$, but the size of $R$ is still bounded by $3r\cdot max\{1,w\}\le \delta/2$, for our setting of parameters. The transformation $\rho^1$ and the shared randomness of the circuits defining the adversary ensure that independently of whether the adversary received a premise $C_0$ or $C_1$ of a cut in $\pi$, the adversary constructs the same $\rho_{\ell}'$ when tracing the corresponding cut in $\pi'$ and responds consistently on the cut in $\pi$. The property that the adversary responds 1 to the axioms of $\neg\PHP^m_n$ is preserved and the probability of responding 0 to the clauses in ${\cal A}$ is $\le (1-\gamma)/2$, if $n$ is big enough.


\def\failreduction{
Given a clause $C$, the adversary sets all variables that do not appear in $\neg\lb^*_W(f_n,n^3/2)$ to 0 and performs a substitution $\rho$ defined as follows:
\begin{itemize}
\item[] $y_{a,j}\mapsto \bigvee_{j'\le j} x_{a,j'}$ if $f(a)=1$; $\neg y_{a,j}\mapsto \bigvee_{j<j'\le s} x_{a,j'}$ if $f(a)=1$; 
\item[] $y^{\epsilon}_{a,j}\mapsto \neg \epsilon$ if $f(a)=0$;
\item[] $s^{\epsilon}_{j,k}\mapsto \bigvee_{\{b \mid b_k=\epsilon\}} x_{b,j}$, for $\epsilon\in \{0,1\}$;
\item[] $y_{a,j,k}\mapsto \bigvee_{\{b \mid a_1=b_1\wedge\dots\wedge a_k=b_k\}} x_{b,j}\vee \bigwedge_{b\in V} \neg x_{b,j}$;
\item[] $\neg y_{a,j,k}\mapsto \bigvee_{\{b \mid a_1\ne b_1\vee\dots\vee a_k\ne b_k\}} x_{b,j}\vee \bigwedge_{b\in V} \neg x_{b,j}$;
\item[] $q_j\mapsto \bigvee_{b\in V} x_{b,j}$.
\end{itemize}

A straightforward verification shows that each axiom of $\neg \lb$ is transformed into a clause which logically follows from $\neg FPHP^{|V|}_s$ (even without using the hole axioms):

\begin{itemize}
\item[] $\neg y_{a,j,k}\vee s^{a_k}_{j,k}\mapsto  \bigvee_{\{b \mid a_1\ne  b_1\vee\dots\vee a_k\ne b_k\}} x_{b,j}\vee  \bigvee_{\{b \mid a_k=b_k\}} x_{b,j}\vee \bigwedge_{b\in V}\neg x_{b,j}$;

\item[] $\neg y_{a,j,k}\vee y_{a,j,k-1}\mapsto  \bigvee_{\{b \mid a_1\ne  b_1\vee\dots\vee a_k\ne b_k\}} x_{b,j}\vee  \bigvee_{\{b \mid a_1=b_1\wedge\dots\wedge a_{k-1}= b_{k-1}\}} x_{b,j}\vee \bigwedge_{b\in V}\neg x_{b,j}$;

\item[] $y_{a,j,k}\vee \neg y_{a,j,k-1}\vee \neg s^{a_k}_{j,k}\mapsto $ 

\quad $\bigvee_{\{b \mid a_1=b_1\wedge\dots \wedge a_k=b_k\}} x_{b,j}\vee  \bigvee_{\{b \mid a_1\ne b_1\vee\dots \vee a_{k-1}\ne b_{k-1}\}} x_{b,j}\vee  \bigvee_{\{b \mid b_k\ne a_k\}} x_{b,j}\vee \bigwedge_{b\in V} \neg x_{b,j}$;

\item[] $\neg y_{a,j}\vee y_{a,j-1}\vee y_{a,j,n} \mapsto \bigvee_{j'>j}  x_{a,j'}\vee \bigvee_{j'\le j-1} x_{a,j'}\vee x_{a,j}\vee\bigwedge_{b\in V} \neg x_{b,j}\vee  \bigvee_{\{b \mid b_k=\epsilon\oplus s_{j,k}\}} x_{b,j} $;

\item[] $y_{a,j}\vee \neg y_{a,j-1} \mapsto \bigvee_{j'\le j}  x_{a,j'}\vee \bigvee_{j'>j-1} x_{a,j'}$;

\item[] $y_{a,j}\vee \neg y_{a,j,n} \mapsto \bigvee_{j'\le j} x_{a,j}\vee \bigvee_{\{b\mid a_1\ne b_1\vee\dots\vee a_k\ne b_k\}} x_{b,j}\vee \bigwedge_{b\in V}\neg x_{b,j}$;

\item[] $y_{a,s}\mapsto \bigvee_{j\le s} x_{a,j}$;
\item[] $\neg y_{a,s}\mapsto 1$.
\end{itemize}

The $\rho$-image of $\pi$ might not be a valid \Res refutation or a valid derivation in any proof system at all. For a clause in $C$, denote by $R(C)$ a set of clauses obtained by expanding $R(q_j)$'s. Note that if $C$ is derived from $C_0,C_1$ in $\pi$ and a partial matching satisfies the $R$-image of $C_0$ and the $R$-image of $C_1$, it satisfies the $R$-image of $C$. Similarly for the weakening rule. In other words, sthe $R$-image of $\pi$ is a valid derivation (in some proof system) w.r.t the assignments respecting the hole axioms.

L3 works for it as well. Now, the proof of the lbs goes through.

One last complication is that the adversary might not know what is the other clause used in the derivation rule. but it knows what it is in the superproof pi'.
}

This completes the proof of Theorem \ref{t:epw}.

\subsection{$k$-evaluations}

One of the most beautiful lower bound methods in proof complexity is the method of $k$-evaluations which was used to prove a lower bound on the length of proofs of the pigeonhole principle $\PHP_n$ (mapping $n+1$ pigeons to $n$ holes) in constant-depth Frege systems \cite{Aphp,KPW,PBI}. 

In the rest of the subsection we prove the following theorem constructing efficient \AC-Frege-adversaries for $\neg \PHP_n$. 

\begin{theorem}[Efficient adversary for \AC-Frege lower bound]\label{t:efadv-keval}\ \newline 
Let $P$ be an $\AC_d$-Frege system in the language consisting of $\vee,\neg,0,1$. Further let $\delta<1/9^d$ and $\gamma<1$. There is $n_0\in\mathbb{N}$ such that for all $n\ge n_0$ there is 
 a $poly(n^{n^{\delta}},m')$-size randomized $P$-adversary $C$ for $\neg \PHP_n$ such that for each $2^{n^{\delta}}$-size $P$-refutation $\pi$ of any set of formulas, with probability $\ge \gamma$ over the randomness of $C$, $\pi$ does not catch $C$. Here $m'$ is the input length of circuits in $C$.
\end{theorem}

We will refer to the presentation of the $\AC_d$-Frege lower bound for $\PHP_n$ in \cite[Chapter 15]{Kpc}.

For a sequence of formulas $\{\theta_i\}$, denote by $\Gamma_{d}\{\theta_i\}$, the set of all depth $\le d$ subformulas of formulas in $\{\theta_i\}$.
\medskip

In each round $j\ge 1$, our adversary will produce a relative $k$-evaluation $(S^{j},H^{j})$ of the set $\Gamma_d\{\phi_{i}|_{\rho'\circ \rho_d}\}_{i\in [j]}$, where $\phi_{i}\in \pi$ is the formula received by the adversary in round $i$ and $k=n^{\delta}$. 
A relative $k$-evaluation is defined as a $k$-evaluation but the base conditions 1 and 2 are dropped. The partial restriction $\rho'$ sets all variables that do not appear in $\PHP_n$ to 0. In order to simplify the notation, we will ignore $\rho'$ in the rest of our proof. The restriction $\rho_d$ will be the restriction $\rho_d$ from the sequence of restrictions $\rho_0\subseteq\dots\subseteq\rho_d$ in the proof of \cite[Lemma 15.2.3]{Kpc} but with $\epsilon_0^d=\epsilon<1/9^d$ instead of $\epsilon<1/5^d$. We assume $\delta<\epsilon$. 
Since the circuits defining the adversary share a random string, the adversary can re-produce the same $\rho_d$ in each round. The response of the adversary to the $j$-th formula $\phi_j$ from a $2^{n^{\delta}}$-size $P$-refutation $\pi$ is 1 if $S^{j}_{\phi_j|_{\rho_d}}(H^{j}_{\phi_j|_{\rho_d}})=S^{j}_{\phi_j|_{\rho_d}}$ and 0 otherwise.

The construction will ensure that there exists a relative $k$-evaluation $(S,H)$ of $\Gamma_d\pi|_{\rho_d}$ such that for each $j$, $(S^j,H^j)$ is $(S,H)$ restricted to $\Gamma_d\{\phi_{i}|_{\rho_d}\}_{i\in [j]}$. In fact, $(S,H)$ will not depend on $\{\phi_i\}_{i\in [j]}$, i.e., independently of whether the adversary traces $\pi$ along the path $\phi_1,\dots,\phi_j$ or along a different path $\phi'_1,\dots,\phi'_j$, it constructs a restriction of a single `global' relative $k$-evaluation $(S,H)$. 

It remains to show how the adversary constructs $(S^{j},H^{j})$ so that they stay consistent w.r.t. all derivation steps in $\pi$ and assign 1 to each clause of $\neg\PHP_n$. 
\medskip

The adversary will construct gradually $(S^{j,t},H^{j,t})$, for each $t\in [d]$, where $(S^{j,t},H^{j,t})$ is a relative $k$-evaluation of $\Gamma_{t}\{\phi_{i}|_{\rho_t}\}_{i\in [j]}$, and set $(S^j,H^j)\colone(S^{j,d},H^{j,d})$. 

If $t$ is the depth of constants $1,0$ or atoms $p_{i,j}$, $(S^{j,t},H^{j,t})$ is an initial (standard) 2-evaluation of subformulas of clauses of $\neg\PHP_n$.

Suppose our adversary already constructed $(S^{j,t-1},H^{j,t-1})$. $(S^{j,t},H^{j,t})$ is obtained from $(S^{j,t-1},H^{j,t-1})$ as follows. First, $\rho_{t}$ is applied on $(S^{j,t-1},H^{j,t-1})$, which by \cite[Lemma 15.2.1]{Kpc} results in a relative $k$-evaluation $(S',H')$ of $\Gamma_{t-1}\{\phi_{i}|_{\rho_t}\}_{i\in [j]}$. Then all formulas in the set $\Gamma_{t}\{\phi_{i}|_{\rho_t}\}_{i\in [j]}\backslash\Gamma_{t-1}\{\phi_i|_{\rho_t}\}_{i\in [j]}$ are gradually `$k$-evaluated' (i.e. added to the domain of $(S',H')$) in an arbitrary order. 
The relative $k$-evaluation obtained at the end of this process is $(S^{j,t},H^{j,t})$. 

If the adversary wants to $k$-evaluate $\neg \psi$ for some $\psi$ that has been $k$-evaluated already, the construction of the corresponding $k$-tree and its subset follows directly from the definition of $k$-evaluation. The only case which requires a more sophisticated treatment is when the adversary needs to $k$-evaluate $\phi\colone\bigvee_i\psi_i$, for $\psi_i$'s that have been already $k$-evaluated. 

In the remaining case, the adversary already possesses $(S'_{\psi_i},H'_{\psi_i})$ given by the relative $k$-evaluation of $\psi_i$'s. The adversary constructs an $n^{O(k)}$-size $k$-tree $S_{\phi}$ such that $\bigcup_{i}H'_{\psi_i}\triangleleft S_{\phi}$ by fixing a strategy $I_{fix}$ on $\bigcup_iH'_{\psi_i}$ and simulating all possible moves of player $II$ in the proof of \cite[Lemma 15.2.2]{Kpc}. The set $H_{\phi}$ is defined as the projection of $\bigcup_i H'_{\psi_i}$ on $S_{\phi}$.

\cite[Lemma 15.2.2]{Kpc}, for $\epsilon<1/9$, does not only show the existence of some $\rho$ such that the desired trees $S_i$ exist, but in fact it shows that these trees exist for a random $\rho$ (chosen from a set of partial matchings) with probability at least $1-(1-\gamma)/d$. In more detail, in the proof of the claim that for some $\rho$, the number of critical pairs is $\le k/2$, we could assume that for $(1-\gamma)/d$ of all $\rho$'s there is a play with at least $k/2$ critical pairs. This would lead to the inequality $(1-\gamma) a\le dsb(2k^{1/2}n^{\epsilon})^k$ which contradicts the setting of the parameters. As a consequence, the probability that the restrictions $\rho_1,\dots,\rho_d$ guarantee that for each relevant $\phi$ the desired $k$-tree $S_{\phi}$ exists 
 is $\ge \gamma$. 

W.l.o.g. the construction of the relative $k$-evaluation of a formula depends only on the formula itself and on $\rho_0\subseteq\dots\subseteq\rho_d$. Since the circuits defining the adversary share randomness, it can be guaranteed that in each round the adversary uses the same $\rho_0,\dots,\rho_d$, independently of how it traces the proof $\pi$, i.e., independently of $\phi_1,\dots,\phi_j$. Therefore, there exists a relative $k$-evaluation $(S,H)$ such that for every path $\phi_1,\dots,\phi_j$ in $\pi$, the adversary constructs $(S,H)$ 
restricted to $\Gamma_d\{\phi_i|_{\rho_d}\}_{i\in [j]}$. 

By \cite[Lemma 15.1.6 and 15.2.1]{Kpc} all clauses of $\neg\PHP_n|_{\rho_d}$ are valid in $(S,H)$, so the adversary responds 1 to the clauses of $\neg\PHP_n$. Since \cite[Lemma 15.1.7]{Kpc} holds even for the relative $k$-evaluation $(S,H)$, which is constructed from the standard 2-evaluation, the adversary will stay consistent w.r.t. all derivation steps.

Finally, generating $(S^{j,t},H^{j,t})$ requires to $k$-evaluate $\le |\pi|^2$ formulas and for each relevant subformula $\phi$, the construction of $(S^{j,t}_{\phi},H^{j,t}_{\phi})$ can be done by a circuit of size $n^{O(k)}$. The complexity of the adversary is thus $poly(n^{n^{\delta}},m')$. This completes the proof of Theorem \ref{t:efadv-keval}.

\section*{AI disclosure} AI was used to (and only to) make several corrections after a preliminary version of the paper appeared in Computational Complexity Conference 2026. Most notably, the definition of pseudo-width in Section \ref{ssec:pwd} was monotonized.

\section*{Acknowledgement}

We are indebted to  Moritz M\"uller for a simplification of the proof of Theorem \ref{t:efadv-inf}, which allowed us to derive Theorem \ref{t:efadv-inf2} for \EF instead of \Gs, and for many comments that clarified the presentation of the paper. We thank Neil Thapen for a correction of Lemma \ref{l:pwlb}, 
Noel Arteche for pointing out to us that Item 2 in Theorem \ref{t:efadv-inf} implies $\NP\not\subseteq\Circuit[n^{\ell}]$, and  Michal Garlik, Pavel Pudl\'ak, Jan Kraj\'i\v{c}ek and anonymous reviewers for suggestions and comments. 
Erfan Khaniki and J\'an Pich received support from the Royal Society University Research Fellowship URF$\backslash$R1$\backslash$211106 ``Proof complexity and circuit complexity: a unified approach". Dmitry Sokolov is supported by the Swiss State Secretariat for Education, Research, and Innovation (SERI) under contract number MB22.00026. This project was initiated at the Institute of Mathematics of the Czech Academy of Sciences. 

\end{document}